\documentclass[dvips,11pt]{article}
\usepackage{amsmath,amssymb,amsfonts,amscd,amsthm,pstricks,pst-node,bm,xspace}
\def\simarrow{\mathrel{\raise -0.5mm\hbox{$\sim$}}\hspace{-1.8mm}{\rightarrow} } 

\def\bsimarrow{\leftarrow\hspace{-0.7mm}\mathrel{\raise -0.5mm\hbox{$\backsim$}} }

\def\bt{\begin{tabular}}
\def\te{\end{tabular}}

\def\lettrine#1#2#3{\noindent\hangindent#1\hangafter-#2
\hskip-#1\smash{\hbox to #1{#3\hfill}}\ignorespaces}

\newcommand{\To}[1]{\mathop{\to}\limits_{#1}}

\def\BM{\begin{pmatrix}}
\def\EM{\end{pmatrix}}

\def\d=f{\buildrel\hbox{\scriptsize d\'{e}f}\over \Longleftrightarrow}

\def\cit{\text{\it I\hskip -6ptC\/}}

\def\rit{\text{\it I\hskip -2pt  R}}

\def\nit{\text{\it I\hskip -2pt  N}}

\def\rl {\rit^{\hskip 1pt\ell}}

\def\Bd{{\text B}}

\def\Ds{{\cal D}}
\def\Ed{{\text E}}

\def\ung{\hbox{1\hskip -4.2pt \rm 1}}

\def\be{\begin{equation}}
\def\ee{\end{equation}}
\def\beqn{\begin{eqnarray}}
\def\eeqn{\end{eqnarray}}
\def\nobeqn{\begin{eqnarray*}}
\def\noeeqn{\end{eqnarray*}}
\def\ba{\left(\begin{array}}
\def\ea{\end{array} \right) }

\def\u{\underline}
\def\o{\overline}
\def\and{\; \mbox{and} \;}

\newcommand{\half}{\frac{1}{2}}

\def\hfl#1#2{\smash{\mathop{\hbox to 12mm{\rightarrowfill}}
\limits^{\scriptstyle #1}_{\scriptstyle #2}}}

\def\mod{\mathop{\rm mod}\nolimits}

\def\Be{\begin{enumerate}}
\def\Ee{\end{enumerate}}

\def\Bena{\begin{enumerate}
\def\labelenumi{\theenumi)}
\def\theenumi{\arabic{enumi}}
\def\labelenumii{\theenumii)}
\def\theenumii{\alph{enumii}}}

\def\Bean{\begin{enumerate}
\def\labelenumii{\theenumii)}
\def\theenumii{\arabic{enumii}}
\def\labelenumi{\theenumi)}
\def\theenumi{\alph{enumi}}}

\def\Bero{\begin{enumerate}
\def\labelenumii{\theenumii)}
\def\theenumii{\arabic{enumii}}
\def\labelenumi{(\theenumi)}
\def\theenumi{\roman{enumi}}}

\def\BeRo{\begin{enumerate}
\def\labelenumii{\theenumii)}
\def\theenumii{\arabic{enumii}}
\def\labelenumi{(\theenumi)}
\def\theenumi{\Roman{enumi}}}

\def\Bi{\vskip 11pt\begin{itemize}\itemsep=18pt}
\def\Ei{\end{itemize}\vskip 11pt}

\def\Bd{\begin{description}}
\def\Ed{\end{description}}

\def\R{\right}
\def\L{\left}

\usepackage[latin1]{inputenc}

\def\st{structure\xspace}

\def\PA{particles\xspace}
\def\bsmpa{bisemiparticle\xspace}
\def\smpa{semiparticle\xspace}
\def\SMPA{semiparticles\xspace}
\def\BSMPA{bisemiparticles\xspace}

\def\bsmst{bisemistructure\xspace}
\def\BSMST{bisemistructures\xspace}
\def\Times{{\mathrel{\u\times}}}
\def\bigoplus{\mathop{\oplus}\limits}

\def\prod{\mathop{\Pi}\limits}
\def\sum{\mathop{\Sigma}\limits}

\def\bbf{\boldmath\bfseries}
\def\o{\overline}

\def\Bi{\begin{itemize}}
\def\Ei{\end{itemize}}

\newcommand{\RR}{\mathbb{R}\,}
\renewcommand{\rit}{\RR}

\def\mod{\operatorname{mod}}

\def\cusp{\operatorname{cusp}}

\def\Rep{\operatorname{Rep}}
\def\Reps{\operatorname{Repsp}}
\def\Repsp{\operatorname{Repsp}}
\def\FRepsp{\operatorname{FRepsp}}

\def\ELLIP{\operatorname{ELLIP}}

\def\GL{\operatorname{GL}}

\def\bM{\begin{matrix}}
\def\eM{\end{matrix}}

\def\lr{left (resp. right) }
\def\rl{right (resp. left) }

\def\resp#1{(resp. #1)}
\def\rresp#1{\qquad \mbox{(resp.} \quad #1\ )}

\def\To{\begin{CD} @>>>\end{CD}}

\def\RL{_{R\times L}}

\def\wt{\widetilde}

\begin{document}

\setcounter{page}{0}
{\pagestyle{empty}
\null\vfill
\begin{center}
{\LARGE Mass generation of elementary particles and origin\linebreak of the fundamental forces in algebraic quantum theory}
\vfill
{\sc C. Pierre\/}
\vskip 11pt


\vfill

\begin{abstract}
The main thesis of this paper deals with the interactions of a set of fermions which are described by:
\Bi
\item one basic type of bilinear interactions;
\item two symmetric semiobjects;
\item three embedded shells;
\item four fundamental (strong) gravito-electro-magnetic forces between left and right semiobjects, i.e. semifermions.
\Ei

\end{abstract}
\vfill
\eject
\end{center}

\vfill\eject}
\setcounter{page}{1}
\def\thepage{\arabic{page}}
\parindent=0pt 
\section{Introduction}

The physics of the twentieth century was based on two great theories: the quantum field theory and the general relativity \cite{Ein}.

{\bf The quantum field theory\/}, in its different versions, describes the quantum fluctuations of the states of the matter and of its interactions in the frame of special relativity.  Its formulation, derived from classical dynamics and relativistic quantum mechanics \cite{Dir}, led us principally up to the existence of antiparticles, to the creation and annihilation of pairs of particles and of quanta from a vacuum state and to an infinite number of degrees of freedom.

But the treatment of interacting fields \cite{Dys} by means of particle exchanges is realized perturbatively which brought divergences probably symptomatic of a chronic disorder in the small distance behavior of the theory.
\vskip 11pt

Quantum field theory  is the framework \cite{Wil} of {\bbf the standard model (SM)\/} \cite{Wei} consisting in a unified description of the strong, weak and electromagnetic forces in the language of quantum Gauge-field theories \cite{A-J}, \cite{A-L}, \cite{G-G-S}.  Despite remarkable successes, the standard model is not expected to be valid at distance scales inferior to $10^{-18}$cm.  It then appears as an excellent low energy approximation of a more fundamental theory\cite{Kot}, \cite{Smo}. Another kind of difficulty of the standard model concerns the {\bbf confinement of the quarks \cite{G-W} in quantum chromodynamics} (QCD).  Indeed, until now, no really quantitative understanding of confinement has been given in the frame of QCD and, furthermore, this low-$Q^2$ region of QCD, relevant to the static properties of baryons, is not perturbatively manageable at large coupling \cite{Kok}.
\vskip 11pt

On the other hand, {\bbf the nonlinear equations of general relativity\/} describe the dynamics of the reciprocical action of the curved space-time on matter in such a way that the geometry of space-time evolves dynamically with respect to the neighbouring external matter.

But, as it is well known, general relativity (GR) cannot be coupled to quantum field theory (QFT) in a standard way because these two theories deal with the description of objects at different length scales.

Only, with {\bf superstring theories (ST)\/}, physicists have succeeded \cite{DelToWit} in incorporating gravity with QFT by introducing massless spin-2 particles in the frame of one-dimensional strings: by this way, the problem of perturbative nonrenormalizability of GR has been solved.  But, the conceptual principles of string theories \cite{G-S-W} have not yet been clarified and, furthermore, these theories have no real contact with experiment \cite{Pol}, \cite{S-S}, \cite{Wit}.
\vskip 11pt

Much of what we know about the standard model and superstring theories comes from perturbation theory described by means of Feynman diagrams of which some are divergent.  These divergences appear as a probable consequence of an over-simple space-time geometry \cite{Ati}.

So, {\bbf a more fundamental theory, taking into account the outstanding features of QFT, SM, GR and ST, should have to give a leading part to the nature of space allowing to consider arbitrary small distance scales\/} \cite{Ash}.  Now, in these theories (QFT, SM and ST), directly deduced from classical dynamics dealing with the study of trajectories of material points, the considered objects (pointlike particles or strings) are quantized by means of commutation or anticommutation relations when {\bf one would expect a pertinent quantization by means of real quanta of structure\/}.  As the space is intrinsically fundamental in physics according to general relativity and in mathematics by the existence of its most famous subjects \cite{Bou} as algebraic geometry,  number theory and algebraic topology, {\bf it would appear reasonable that a quantum theory in physics be quantized ``ab initio'' by a quantization of space\/} whose quanta would have a well defined mathematical structure \cite{Pie4}.
\vskip 11pt

This challenge was achieved in {\bbf the algebraic quantum theory (AQT) which is a quantum theory devoted to the elementary particle internal structure based on the quantization of their internal space(-time) by algebraic quanta being algebraic closed subsets of degree $N$\/}.
This internal space structure is then discrete and, under some (toroidal) compactification, gives rise to a continue geometry and to analytic functions (or cuspidal forms) by means of the procedure of the Langlands global program \cite{Pie2}, \cite{Car}, \cite{Har}, \cite{Lan1}, \cite{Lan2}, \cite{Gel}. 
By this way, the algebraic quantum theory, resulting from a unique conceptual framework, gives rise to physical quantities and structures which are all quantized in the frame of algebraic geometry and analytic number theory and which refer to the pertinent features of the quantum field theories, the general relativity, the standard model and the string theories.
\vskip 11pt

More concretely, the algebraic quantum theory proceeds form {\bf the unification of quantum field theory with general relativity} as proved in \cite{Pie3} which {\bf implies that elementary particles have to be endowed with an internal structure composed of three embedded substructures labeled ST, MG and M (for space-time, middle ground and mass)} in such a way that the mass shell structure ``$M$'' is generated (or created) from two most internal shell substructures ST and MG constituting the particle internal vacuum.  {\bf The origin and the nature of the mass of an elementary particle is then clearly identified mathematically as resulting from its internal vacuum being of space-time nature\/}.
\vskip 11pt

On the other hand, {\bf QFT, implying its fusion with special relativity (SR),} leads to the existence of antiparticles, directly related to the bilinear invariants of SR.  This was transposed in AQT by developing a theory of bilinear structure of elementary particles which are then envisaged as bisemiparticles: a bisemiparticle is defined by the product of a left semiparticle, localized in the upper half space, by its symmetric right equivalent localized in the lower half space.
\vskip 11pt

{\bf A bisemiparticle is thus a mathematical  bisemiobject characterized by a bisemistructure \cite{Pie6}, resulting from the reciprocal action of the left semistructure on the right semistructure\/}. The bielements of this bisemistructure are either diagonal bielements referring to the central diagonal bisemistructure of a bisemiparticle or cross bielements responsible for the off-diagonal magnetic and electric structures of the bisemiparticle.
{\bf An elementary fermion can then be viewed as an elementary bisemifermion described by the product of a left semifermion by its  symmetric right equivalent\/} in such a way that, under some external perturbation, this bisemifermion could be split generating a pair of fermion-antifermion of which fermion corresponds to the left semifermion and antifermion to the right semifermion.  {\bf In a bisemifermion, the right semifermion, corresponding to the antifermion, is projected onto the associated left semifermion and is hidden by the only observable (left semi)fermion} \cite{Pie1}: {\bf these hidden right semifermions could correspond partly to the dark matter.}

{\bbf In special relativity, time has been linked with space into a four-dimensional space-time.  But,  time can be really transformed into space and vice-versa   if time has the same type of topological structure as space at the subplanckian quantum level.\/}  So, a discrete structural quantum time has been envisaged as constituting the time part of the space-time (ST) internal vacuum of every elementary (bisemi)fermion and the ratio of the space structure by the time structure, generated from it and localized in an orthogonal space, is given by the quotient between the two corresponding Hecke parameters \cite{Pie1}.
This ratio of space internal structure by the time internal structure gives the velocity of the shell structure ``~ST~'' of the elementary (bisemi)fermion and corresponds to the velocity of light appearing in the equations of QFT when the time structure tends to zero.  This brings {\bf a new ontological meaning to the concept of velocity} (of light) {\bf at the level of elementary particles\/}.
\vskip 11pt

The central notion of field of QFT is defined mathematically in AQT by means of the global program of Langlands in the case $\GL(2)$.  Indeed, it was proved in \cite{Pie4} that {\bf a time} (or space) {\bf string field of the internal vacuum of an elementary bisemifermion} of the first family is given by the bisemisheaf $(\wt M^T_{{\rm ST}_R} \otimes_{(D)} \wt M_{{\rm ST}_L})$ of a $\cit$-valued differentiable bifunctions 
$(\phi _R(M_{L_{\o v_{\mu ,m_\mu }}})\otimes_{(D)}
\phi _L(M_{L_{v_{\mu ,m_\mu }}}))$ on the set of conjugacy class representatives
$(M_{L_{\o v_{\mu ,m_\mu }}}\otimes_{(D)}
M_{L_{v_{\mu ,m_\mu }}})$, $1\le \mu \le q\le\infty $, of the bilinear algebraic semigroup $\GL_2(L_{\o v}\times L_v)$ over the product $(L_{\o v}\times L_v)$ of the sets of completions $L_{\o v}$ and $L_v$ \cite{Pie5}.

Each differentiable bifunction, i.e. the product of a right function localized in the lower half space by its symmetric left equivalent localized in the upper half space, of the bisemisheaf
$(\wt M^T_{{\rm ST}_R}\otimes_{(D)} \wt M^T_{{\rm ST}_L})$ is a bistring, i.e. the product of a right string by a left symmetric string, in such a way that every right and left string is characterized by a structure at $\mu $ quanta.

{\bbf Thus a bisemisheaf $(\wt M^T_{{\rm ST}_R}\otimes_{(D)} \wt M^T_{{\rm ST}_L})$ is a time (or space) string field because it is composed of a tower of increasing bistrings behaving like harmonic oscillators and characterized by a number of increasing biquanta\/}.

As above mentioned, {\bbf a quantum is an algebraic closed subset of degree $N$\/}.  If we are dealing with a quantum localized on the mass shell, then the Planck constant $\hbar$ must correspond to the number $N$ in the mass system of units \cite{Pie1}, giving then a mathematical origin to $\hbar$.

An important feature of QFT concerns the creation and annihilation of (bi)quanta on (bi)strings; this is realized in AQT respectively by Galois automorphisms \cite{Rib} and antiautomorphisms corresponding to quantum deformations and inverse quantum deformations of these bistrings \cite{Pie1}.
\vskip 11pt

As a consequence, {\bbf the considered time (or space) string field is a bosonic string field}
which is such that bistrings can be added (i.e. created) or removed (i.e. annihilated) to or from this field by Galois automorphisms or antiautomorphisms.  If this bosonic string field is an operator valued string field \cite{Pie5}, {\bbf then the spectrum of the considered operator acting on the bosonic string field is naturally generated by envisaging creation (and/or annihilation) of bistrings\/} \cite{Pie1}.

All that concerns {\bf elementary (bisemi)fermions}, i.e. the leptons $e^-$, $\mu ^-$, $\tau ^-$ and their neutrinos as well as the quarks $u^+$, $d^-$, $s^-$, $c^+$, $b^-$ and $t^+$.

{\bbf The composite (bisemi)fermions are the baryons} which are described in the frame of quantum chromodynamics by hadronic bound states confining three colored quarks.  But, {\bf the origin of this confinement} still remains mysterious.  This problem {\bf was reconsidered in AQT in such a way that the three \lr semiquarks of a bisemibaryon originate and are connected to a \lr (massive) ``core'' time semifield of the \lr semibaryon\/}.  {\bbf The confinement of the (semi)quarks then arises naturally and  the strong interaction depends on the existence of a strong gravitational and electric field resulting from the bilinear interactions between the central core structures of the left and right semibaryons and the structures of the right and left semiquarks\/}.  The strong constant is then directly related to the generation of the three semiquarks from the central core structure of the corresponding semibaryon as it was proved in \cite{Pie1}.
\vskip 11pt

All that leads to the {\bbf main proposition} of this paper:
\vskip 11pt

\paragraph{Proposition:} \mbox{}

{\em 
{\bbf The bisemifermions (i.e. the elementary (bisemi)fermions and the (bisemi)baryons) are defined by means of:
\Bi
\item $1$ basic type of bilinear interaction;
\item $2$ symmetric semiobjects (which are semiparticles);
\item $3$ internal embedded shells: ST, MG and M~;
\item $4$ fundamental forces at the level of each shell ST~, MG and M plus 4 mixed fundamental forces on products, right by left, of different shells ST $\times$ MG~, ST $\times$ M and MG  $\times$ M~;
\Ei
}
in such a way that: the set of the symmetric left and right semifermions, localized respectively in the upper and lower half spaces and characterized by three embedded substructures ST~, MG and M~, interacts by means of a general type of bilinear interaction which is of gravito-electro-magnetic (GEM) type.

{\bbf This basic type of ``GEM'' bilinear interaction between interacting right and left semifermions is at the origin of:
\Bean 
\item the gravitational and electromagnetic forces mediated by massless GEM bisemi\-bosons;
\item the strong and weak forces mediated by GEM massive (bisemi)bosons.
\Ee}}
\vskip 11pt

In order to prove this proposition, the chapter 2 of this paper will be devoted to the study of the internal structure of elementary (bisemi)particles leading to a revised classification of these intro (bisemi)fermions and into (bisemi)bosons obeying respectively the Fermi-Dirac and the Bose-Einstein statistics.  The spin, directly related to the rotation of the internal structure of particles, then received a new ontological meaning.

In chapter 3, the basic types of gravito-electro-magnetic bilinear interactions leading to the four fundamental forces of physics as well as to the three new forces between the internal shells of particles is studied.

Finally, it may be remarked that bringing to light the internal structure of elementary particles in AQT allows to precise {\bf what are really the mass, the electric charge, the spin, the nature of space-time and the gravitation.}
\vskip 11pt
\section{Internal quantum structure of \BSMPA}

In this chapter, the \st of  elementary \PA, introduced in AQT \cite{Pie1}, will be reviewed and succinctly justified.
\vskip 11pt

\subsection{Justifications of an internal bilinear structure for elementary particles}

\Bi
\item To take really into account the bilinear invariants of special relativity, {\bbf \PA will be described as \BSMPA\/}, i.e. by the (bilinear tensor) product of a left semiparticle, localized in the upper half space, by its symmetric right equivalent, localized in the lower half space.
\vskip 11pt

\item The unification of quantum field theory with general relativity \cite{Pie3} implies in fact that every elementary (bisemi)particle be endowed with {\bbf an internal space-time (bisemi)\-structure constituting its own vacuum from which its mass shell can be generated\/} \cite{Pie3}.
\vskip 11pt

\item More concretely, {\bbf every bisemifermion is then composed of three embedded (bi\-semi)structures ST $\subset$ MG $\subset$ M which are  its internal fields\/}.  Each \bsmst ST~, MG or M splits into a diagonal central \bsmst and into two off-diagonal magnetic and electric \BSMST according to the definition of a bilinear semigroup \cite{Pie6}.
\vskip 11pt

\item This is the case for the {\bbf simple bisemifermions} of the first family, i.e. the electron $e^-$ (and its neutrino $\nu _{e^-}$) and the quarks $u^+$ and $d^-$, but also for the simple bisemifermions of the second and third families.
The (bisemi)baryons which are {\bbf composite bisemifermions\/} are characterized in the frame of AQT by a central core time \bsmst ST $\cup$ MG $\cup$ M to which the (bisemi)quarks are connected as it will be recalled in the following.
\vskip 11pt

\item One of the main aims of the algebraic quantum theory consists in {\bbf quantizing the internal fields ST~, MG and M of bisemiparticles  by (bi)quanta of structure} allowing quantum jumps.  This implies that this theory must be algebraic or, more exactly, algebraico-geometric in such a way that the quanta {\bf have a special structure given by algebraic irreducible closed subsets} characterized by a Galois extension degree $N$.
\vskip 11pt

\item {\bbf Each internal physical field ST~, MG or M refers to a correspondence of the Langlands global program} since it is a (bisemi)sheaf of differentiable (bi)functions on a bilinear algebraic semigroup $\GL_2(L_{\o v}\times L_v)$ (as mentioned in the introduction) in such a way that this (bisemi)sheaf constitutes the $2$-dimensional holomorphic (or cuspidal) representation of the considered Weil (bisemi)group.
\Ei
\vskip 11pt

\subsection{Quantum structure of internal fields}

{\bf The internal fields} of an elementary particle {\bf must}:
\Bean
\item {\bf have a primary discrete character} in order to be quantized by algebraic quanta of structure \cite{Pie4};

\item but, also, {\bf have a secondary continue character}  leading to a holomorphic function or a cuspical form;
\item {\bf be graded} in function of the number of quanta in order to give rise to quantum jumps;
\item {\bf be bilinear} as mentioned.
\Ee
\vskip 11pt

This challenge can only be reached in the context of the Langlands global program by considering the representations of algebraic bilinear semigroups.
\vskip 11pt

Let then $K$ be a global number field of characteristic $0$ and let $F_{\o v}\cup F_v$ denote the real algebraic symmetric splitting field \cite{Pie1}, \cite{Pie4}, over $K$ built from an increasing number of irreducible closed real algebraic subsets $F_{\o v^1}\cup F_{v^1}$ where
$F_{v^1}$ refers to a positive symmetric splitting subsemifield and 
$F_{\o v^1}$  to its negative symmetric equivalent.

Each irreducible closed real algebraic subset $F_{v^1}$ or $F_{\o v^1}$ is characterized by a Galois extension degree 
$[F_{v^1}:K]=[F_{\o v^1}:K]=N$, $N\in\nit$, and is interpreted as {\bbf a quantum of structure which can be of time or space type.}

Let $F_v=\{F_{v_1},\dots,F_{v_{\mu ,m_\mu }},\dots,F_{v_{q,m_q}}\}$
\resp{$F_{\o v}=\{F_{\o v_1},\dots,F_{\o v_{\mu ,m_\mu }},\dots,F_{\o v_{q,m_q}}\}$} denote the set of real algebraic subsets (or symmetric splitting semifields) $F_{v_\mu }$ \resp{$F_{\o v_\mu }$} characterized by their Galois extension degrees
\[ [F_{v_\mu }:K]\equiv[F_{\o v_\mu }:K]
=*+\mu N\;, \qquad 1\le \mu \le q\le \infty \;, \]
which are integers modulo $N$, where $*$ denotes an integer inferior to $N$.

So, we get a tower
\[
F_{v_1}  \subset \dots \subset F_{v_{\mu ,m_\mu }} \subset \dots \subset F_{v_{q ,m_q }}\qquad
\rresp{F_{\o v_1}  \subset \dots \subset F_{\o v_{\mu ,m_\mu }} \subset \dots \subset F_{\o v_{q ,m_q }}}
\]
of classes of real algebraic subsets in such a way that the $m_\mu $-th representative of the $\mu $-th equivalence class has a structure at $\mu $ quanta of degree $N$.

The Weil group is assumed to be the Galois subgroup of the extensions characterized by degrees $d=0\mod N$ \cite{Pie2}.
\vskip 11pt

\subsection{Proposition}
{\em
{\bbf Let $\GL_2(F_{\o v}\times F_v)\equiv T^t_2(F_{\o v})\times T_2(F_v)$ be the algebraic bilinear semigroup} of which representation space $\Repsp(\GL_2(F_{\o v}\times F_v))$ is the tensor product 
$M_R(F_{\o v})\otimes M_L(F_v)$ of a right
$T^t_2(F_{\o v})$-semimodule $M_R(F_{\o v})$ by the symmetric left
$T _2(F_{v})$-semimodule $M_L(F_{v})$.

Then, the $\GL_2(F_{\o v}\times F_v)$-bisemimodule
$M_R(F_{\o v})\otimes M_L(F_v)$ splits into:
\[
M_R(F_{\o v})\otimes M_L(F_v)\simeq
(M_R(F_{\o v})\otimes_D M_L(F_v))\oplus
(M_R(F_{\o v})\otimes_m M_L(F_v))\]
where:
\Bi
\item $M_R(F_{\o v})\otimes_D M_L(F_v)$ is a diagonal tensor product;
\item $M_R(F_{\o v})\otimes_m M_L(F_v)$ refers to an off-diagonal tensor product of magnetic type.
\Ei
}
\vskip 11pt

\begin{proof}
\Bena
\item The algebraic bilinear semigroup
$\GL_2(F_{\o v}\times F_v)$
was introduced in order to reflect the bilinear character of the matrices.  It was shown in \cite{Pie2} that the algebraic bilinear semigroup 
$\GL_2(F_{\o v}\times F_v)$ covers its linear equivalent
$\GL_2(F_{\o v}- F_v)$ where $F_{\o v}- F_v$
means $F_{\o v}\cup F_v$.

\item As the product $( F_{\o v}\times F_v)$ is taken over corresponding pairs
$\{F_{\o v_{\mu ,m_\mu }},F_{v_{\mu ,m_\mu }}\}$ of right and left algebraic subsets,
$\Repsp \GL_2(F_{\o v}\times F_v)$ decomposes into the set
$\{\GL_2(F_{\o v_{\mu ,m_\mu }}\times F_{v_{\mu ,m_\mu }})\}_{\mu ,m_\mu }$ of conjugacy class representatives of
$\GL_2(F_{\o v}\times F_v)$.

\item As $\GL_2(F_{\o v}\times F_v)$ is a bilinear semigroup, its representation space
$M_R(F_{\o v})\otimes M_L(F_v)$, being a 
$\GL_2(F_{\o v}\times F_v)$-bisemimodule, splits according to \cite{Pie6}:
\begin{align*}
M_R(F_{\o v})\otimes M_L(F_v)
&\simeq (M_R(F_{\o v})\otimes_D M_L(F_v))\oplus
 (M_R(F_{\o v})\otimes_m M_L(F_v))\\[8pt]
 &\simeq \{
 M_{F_{\o v_{\mu ,m_\mu }}}\otimes_D
 M_{F_{v_{\mu ,m_\mu }}}\}_{\mu ,m_\mu }
 \oplus
 \{
 M_{F_{\o v_{\mu ,m_\mu }}}\otimes_m
 M_{F_{v_{\mu ,m_\mu }}}\}_{\mu ,m_\mu }
 \end{align*}
 where $
 M_{F_{v_{\mu ,m_\mu }}}$
 \resp{$M_{F_{\o v_{\mu ,m_\mu }}}$}
 is the $(\mu ,m_\mu )$-th representation subspace of the \lr conjugacy class
$T_2(F_{v_{\mu ,m_\mu }})$
\resp{$T^t_2(F_{\o v_{\mu ,m_\mu }})$} restricted to the upper \resp{lower} half space \cite{Mum}.\qedhere
\Ee
\end{proof}
\vskip 11pt

\subsection{Proposition}

{\bfseries {\em The splitting of the representation space
{\bbf $(M_R(F_{\o v})\otimes M_L(F_v))$} of the bilinear algebraic semigroup {\bbf $\GL_2(F_{\o v}\times F_v)$}}} {\em into a diagonal and a magnetic representation subspace  $(M_R(F_{\o v})\otimes_D M_L(F_v))$ and
$(M_R(F_{\o v})\otimes_m M_L(F_v))$
{\bfseries implies an internal dynamics of
$(M_R(F_{\o v})\otimes M_L(F_v))$ given by a rotation in opposite directions of the left and right representation subspaces
$M_{F_{v_{\mu ,m_\mu }}}$ and
$M_{F_{\o v_{\mu ,m_\mu }}}$}.

As a consequence, {\bfseries these representation subspaces are compactified and cuspidal representations} (as well as holomorphic functions)
{\bfseries can be defined on them by means of Langlands global correspondences} \cite{Pie2}}.
\vskip 11pt

\begin{proof}
\Bena
\item The splitting of the representation space of
$\GL_2(F_{\o v}\times F_v)$ into:
\[ M_R(F_{\o v})\otimes M_L(F_v)=
 (M_R(F_{\o v})\otimes_D M_L(F_v))+
 (M_R(F_{\o v})\otimes_m M_L(F_v))\]
 corresponds to an exchange of magnetic biquanta between the right and left conjugacy class representatives
 $M_{F_{\o v_{\mu ,m_\mu }}}\subset M_R(F_{\o v})$ and
 $M_{F_{v_{\mu ,m_\mu }}}\subset M_L(F_{v})$
 as it was described in \cite{Pie5} by means of Feynman diagrams.
 
 Now, this exchange of magnetic biquanta implies an internal dynamics of these left and right conjugacy class representatives appearing by their rotation in opposite senses in such a way that different magnitudes of rotational velocities in each pair
$\{M_{F_{\o v_{\mu ,m_\mu }}},M_{F_{v_{\mu ,m_\mu }}}\}$ lead to an exchange of magnetic biquanta in it, as proved in \cite{Pie1}.

\item The rotation of the ``one-dimensional'' conjugacy class representatives
$M_{F_{\o v_{\mu ,m_\mu }}}$ and\linebreak
 $M_{F_{v_{\mu ,m_\mu }}}$, at the origin of the spin of
 $M_R(F_{\o v})$ and $M_L(F_v)$ respectively, leads to a compactification 
 $\gamma ^c_{M_{F_{\o v_{\mu ,m_\mu }}}}$
 \resp{$\gamma ^c_{M_{F_{v_{\mu ,m_\mu }}}}$} of them by:
 \begin{align*}
\gamma ^c_{M_{F_{v_{\mu ,m_\mu }}}}: \quad
M_{F_{v_{\mu ,m_\mu }}}&\To M_{L_{v_{\mu ,m_\mu }}}
\\[8pt]
\rresp{\gamma ^c_{M_{F_{\o v_{\mu ,m_\mu }}}}: \quad
M_{F_{\o v_{\mu ,m_\mu }}}&\To M_{L_{\o v_{\mu ,m_\mu }}}}\end{align*}
as developed in \cite{Pie3}.

By this way, the algebraic subsets
$M_{F_{v_{\mu ,m_\mu }}}$
\resp{$M_{F_{\o v_{\mu ,m_\mu }}}$}
are transformed into the completions
$M_{L_{v_{\mu ,m_\mu }}}$
\resp{$M_{L_{\o v_{\mu ,m_\mu }}}$}
at $\mu $ quanta viewed as quantum ``big points'' and
$\Repsp(\GL_2(F_{\o v}\times F_v))$ is transformed into
$\Repsp(\GL_2(L_{\o v}\times L_v))=
M_R(L_{\o v})\otimes M_L(L_v))$.

\item On each conjugacy class representation subspace
$M_{L_{v_{\mu ,m_\mu }}}\subset M_{L_{\omega _\mu}} $
\resp{$M_{L_{\o v_{\mu ,m_\mu }}}\subset M_{L_{\o\omega _\mu }}$} included in its complex equivalent
$M_{L_{\omega _\mu }}$
\resp{$M_{L_{\o\omega _\mu }}$}, we can define the function
\[ f_{v_\mu }(z^\mu ): \quad M_{L_{v_{\mu ,m_\mu }}}\To\cit \qquad
\rresp{f_{\o v_\mu }(z^{*\mu }): \quad M_{L_{\o v_{\mu ,m_\mu }}}\To\cit}\]
in such a way that, on the set of compactified representation subspaces
$M_{L_{v_{\mu ,m_\mu }}}$
\resp{$M_{L_{\o v_{\mu ,m_\mu }}}$}, the function:
\begin{align*}
f_v(z) = \sum_\mu  \sum_{m_\mu } c_{\mu ,m_\mu }(z-z_0)^\mu 
\rresp{f_{\o v}(z^*) = \sum_\mu  \sum_{m_\mu } c^*_{\mu ,m_\mu }(z^*-z^*_0)^\mu },\end{align*}
defined in a neighbourhood of a point $z_0$ \resp{$z^*_0$} of $\cit$, is holomorphic at $z_0$ \resp{$z^*_0$} where $z$ \resp{$z^*$} is a complex function of one real variable \cite{Pie2}.

\item Similarly, if the compactification of the conjugacy class representatives
$M_{F_{v_{\mu ,m_\mu }}}$
\resp{$M_{F_{\o v_{\mu ,m_\mu }}}$} is a toroidal compactification mapping them  into the circles
$M^T_{L_{v_{\mu ,m_\mu }}}$
\resp{$M^T_{L_{\o v_{\mu ,m_\mu }}}$}, then we can define on them the circular functions:

\begin{align*}
g_\mu (e^{2\pi i\mu x}): \qquad M^T_{L_{v_{\mu ,m_\mu }}}&\To \cit\;, \qquad x\in\rit\;, \\[8pt]
\rresp{g_\mu (e^{-2\pi i\mu x}): \qquad M^T_{L_{\o v_{\mu ,m_\mu }}}&\To \cit}\end{align*}
in such a way that their sum generates the (truncated) global elliptic semimodule:
\begin{align*}
\ELLIP_L(1,\mu ,m_\mu )&= \sum_{\mu =1}^q \sum_{m_\mu } \lambda ^{\half}(1,\mu ,m_\mu )\ e^{2\pi i\mu x}\\[8pt]
\rresp{\ELLIP_R(1,\mu ,m_\mu )&= \sum_{\mu =1}^q \sum_{m_\mu } \lambda ^{\half}(1,\mu ,m_\mu )\ e^{-2\pi i\mu x}}
\end{align*}
where $\lambda (1,\mu ,m_\mu )$ is the eigenvalue of the $(\mu ,m_\mu )$-th coset representative of the product, right by left, of Hecke operators \cite{Pie2}.

\item Finally, it was proved that
$(\ELLIP_R(1,\mu ,m_\mu )\otimes_{(D)}
\ELLIP_L(1,\mu ,m_\mu ))$ constitutes an irreducible cuspidal representation \cite{Lang} of the bilinear algebraic semigroup
$\GL_2(F_{\o v}\times F_v)$ and that
$\GL_2(F_{\o v}\times F_v)$ is the representation of the product, right by left, of the considered global Weil (semi)groups \cite{Pie7}.

This leads to the Langlands global correspondence:
\[ \Reps\cusp_{\GL_2}:\qquad
\GL_2(F_{\o v}\times F_v) \To
\ELLIP_R(1,\mu ,m_\mu ) \otimes_{(D)}
\ELLIP_L(1,\mu ,m_\mu )\;.\qedhere\]
\Ee
\end{proof}

\subsection[Bisemisheaf of differentiable bifunctions on \bbf 
$M_{L_{\o v_{\mu ,m_\mu }}}\otimes
M_{L_{v_{\mu ,m_\mu }}}$]{\bbf Bisemisheaf of differentiable bifunctions on 
$M_{L_{\o v_{\mu ,m_\mu }}}\otimes
M_{L_{v_{\mu ,m_\mu }}}$}

\Bi
\item The set $\{\phi _L(M_{L_{v_{\mu ,m_\mu }}})\}_{\mu ,m_\mu }$
\resp{ $\{\phi _R(M_{L_{\o v_{\mu ,m_\mu }}})\}_{\mu ,m_\mu }$} of $\cit$-valued differentiable functions on the compactified conjugacy class representation subspaces
$M_{L_{v_{\mu ,m_\mu }}}$
\resp{\linebreak $M_{L_{\o v_{\mu ,m_\mu }}}$} constitutes the set 
$\Gamma (\wt M^T_{\rm ST_L})$
\resp{$\Gamma (\wt M^T_{\rm ST_R})$} of sections of the semisheaf of rings
$\wt M^T_{\rm ST_L}$
\resp{$\wt M^T_{\rm ST_R}$}.
\vskip 11pt

\item The set 
$\{\phi _R(M_{L_{\o v_{\mu ,m_\mu }}})\otimes
\phi _L(M_{L_{v_{\mu ,m_\mu }}})\}_{\mu ,m_\mu }$
of differentiable bifunctions over the $\GL_2(L_{\o v}\times L_v)$-bisemimodule
$M_R(L_{\o v})\otimes
M_L(L_{v})$, also written
$( M^T_{\rm ST_R}\otimes
M^T_{\rm ST_L})$ if it refers to the ``time'' bisemisheaf, is the set of bisections of the bisemisheaf of rings
$(\wt M^T_{\rm ST_R}\otimes
\wt M^T_{\rm ST_L})$.
\Ei
\vskip 11pt

\subsection{Proposition}

{\em
{\bbf The bisemisheaf $(\wt M^T_{\rm ST_R}\otimes
\wt M^T_{\rm ST_L})$ of differentiable bifunctions is a physical bosonic time string field} of the internal vacuum of a simple bisemifermion.

This bisemisheaf is {\bbf characterized by:}
\Bean
\item {\bbf the number of its bisections which are bistrings;}
\item {\bbf the number of quanta on each pair of symmetric left or right section (or string).}
\Ee}
\vskip 11pt

\begin{proof}
\Bena
\item The bisemisheaf
$(\wt M^T_{\rm ST_R}\otimes
\wt M^T_{\rm ST_L})$ is a physical string field because each product, right by left,
$(\phi _R(M_{L_{\o v_{\mu ,m_\mu }}})\otimes
\phi _L(M_{L_{v_{\mu ,m_\mu }}}))$ of corresponding sections, which are one-dimensional, is a bistring behaving like a harmonic oscillator as proved in \cite{Pie1}.

\item Each left or right string of a bistring
$(\phi _R(M_{L_{\o v_{\mu ,m_\mu }}})\otimes
\phi _L(M_{L_{v_{\mu ,m_\mu }}}))$ is composed of $\mu $ quanta of structure according to section 2.2.

If the string field
$(\wt M^T_{\rm ST_R}\otimes
\wt M^T_{\rm ST_L})$ is composed of $q$ bisections or bistrings, $1\le q\le \infty $,  the integer $q$ is interpreted as its number of degrees of freedom or normal modes \cite{Vneu}.

\item The string field $(\wt M^T_{\rm ST_R}\otimes
\wt M^T_{\rm ST_L})$ is a bosonic field because each bisection $\phi _R(M_{L_{\o v_{\mu ,m_\mu }}})\otimes
\phi _L(M_{L_{v_{\mu ,m_\mu }}})$ is a bisemiboson (i.e. the product of a right semiboson localized in the lower half space by its symmetric left equivalent localized in the upper half space) embedded in $(\wt M^T_{\rm ST_R}\otimes
\wt M^T_{\rm ST_L})$.

Indeed, each bisection at $\mu $ biquanta, written in condensed form $(\phi _{\mu _R}\otimes \phi _{\mu _L})$, is submitted to {\bbf quantization rules consisting in} \cite{Pie1}:
\Be
\item adding $\nu $ biquanta to it, which corresponds to the quantum deformation:
\[ \Ds\RL^{[\mu ]\to[\mu +\nu ]}: \qquad 
\phi _{\mu _R}\otimes \phi _{\mu _L}\To
\phi _{(\mu+\nu ) _R}\otimes \phi _{(\mu +\nu )_L}\]
associated to the exact sequence:
\[ \ung \To \wt M\RL^I 
\To \phi _{(\mu+\nu ) _R}\otimes \phi _{(\mu+\nu ) _L}
\To \phi _{\mu  _R}\otimes \phi _{\mu  _L}\To\ung\]
of which kernel $\wt M\RL^I $ is a biquantum.

\item or removing $\nu $ biquanta from it which corresponds to the inverse quantum deformation:
\[ \Ds\RL^{[\mu +\nu]\to[\mu  ]}: \qquad 
\phi _{(\mu+\nu ) _R}\otimes \phi _{(\mu +\nu )_L}\To
\phi _{\mu _R}\otimes \phi _{\mu _L}
\]
in such a way that $\nu $ biquanta become free.
\Ee

The quantum deformation $\Ds\RL^{[\mu ]\to[\mu +\nu ]}$ results from Galois automorphisms while the inverse quantum deformation $\Ds\RL^{[\mu +\nu]\to[\mu  ]}$ corresponds to an endomorphism based on Galois antiautomorphisms \cite{Pie1}.

Remark that the quantum deformation
$\Ds\RL^{[\mu ]\to[\mu +\nu ]}$ 
of the $\mu $-th bisection
$(\phi _{\mu _R}\otimes\phi _{\mu _L})$ of a physical field corresponds to a local bilinear Gauge transformation of $(\phi _{\mu _R}\otimes\phi _{\mu _L})$ \cite{Pie1}.\qedhere
\Ee
\end{proof}
\vskip 11pt

\subsection{Corollary}

{\em {\bbf The (time) bosonic string field
$(\wt M^T_{\rm ST_R}\otimes
\wt M^T_{\rm ST_L})$ has a cuspidal representation} given by:
\[ \Rep\cusp_{\wt M^T_{\rm ST_R}\otimes
\wt M^T_{\rm ST_L}}: \qquad
\wt M^T_{\rm ST_R}\otimes
\wt M^T_{\rm ST_L} \To
\ELLIP_R(1,\mu ,m_\mu )\otimes_{(D)}
\ELLIP_L(1,\mu ,m_\mu )\]
where $\ELLIP_R(1,\mu ,m_\mu )\otimes_{(D)}
\ELLIP_L(1,\mu ,m_\mu )$ is the global elliptic bisemimodule introduced in proposition 2.4.}
\vskip 11pt

\begin{proof}
This is evident since the bisemisheaf
$(\wt M^T_{\rm ST_R}\otimes
\wt M^T_{\rm ST_L})$ is a functional representation space of the bilinear algebraic semigroup
$\GL_2(F_{\o v}\times F_v)$ according to section 2.5.
\end{proof}
\vskip 11pt

\subsection{Operator-valued string field}

{\bbf The operator-valued string field
$(\wt M^{T_p}_{\rm ST_R}\otimes
\wt M^{T_p}_{\rm ST_L})$} is obtained  from
$(\wt M^T_{\rm ST_R}\otimes
\wt M^T_{\rm ST_L})$ by action of the differential bioperator:
\[ T_{R;{\rm ST}}\otimes T_{L;{\rm ST}}: \qquad
\phi _{\mu _R}\otimes \phi _{\mu _L} \To
\L(-i\ \dfrac{\hbar_{\rm ST}}{c_{t\to r;\rm ST}}\ 
\vec s_{0_R} \dfrac\partial{\partial t_0}\ \phi _{\mu _R}\R)\otimes
\L(i\ \dfrac{\hbar_{\rm ST}}{c_{t\to r;\rm ST}}\ 
\vec s_{0_L} \dfrac\partial{\partial t_0}\ \phi _{\mu _L}\R)\]
on each of its bisections $(\phi _{\mu _R}\otimes\phi _{\mu _L})\in
(\wt M^T_{\rm ST_R}\otimes
\wt M^T_{\rm ST_L})$, $1\le \mu \le q$, where:
\Bi
\item $\hbar_{\rm ST}$ and $c_{t\to r}$ are respectively the Planck constant and the velocity of light at this (space-) time internal level \cite{Pie1};

\item $\vec s_{0_L} \dfrac\partial{\partial t_0}\ \phi _{\mu _L}$
\resp{$\vec s_{0_R} \dfrac\partial{\partial t_0}\ \phi _{\mu _R}$} denotes the directional derivative of $\phi _{\mu _L}$ \resp{$\phi _{\mu _R}$} in the time variable $t_0$ in the direction 
$\vec s_{0_L}$
\resp{$\vec s_{0_R}$}.
\Ei
So, we have that:
\[ T_{R;\rm ST}\otimes T_{L;\rm ST}: \qquad
(\wt M^{T}_{\rm ST_R}\otimes
\wt M^{T}_{\rm ST_L})\To
(\wt M^{T_p}_{\rm ST_R}\otimes
\wt M^{T_p}_{\rm ST_L})\;.\]
\vskip 11pt

\subsection{Proposition}

{\em   {\bbf The operator-valued string field
$(\wt M^{T_p}_{\rm ST_R}\otimes
\wt M^{T_p}_{\rm ST_L})$}, total bisemispace of the tangent bibundle
$(\wt M^{T_p}_{\rm ST_R}\otimes
\wt M^{T_p}_{\rm ST_L},
(T_{R;{\rm ST}}\otimes T_{L;{\rm ST}})^{-1},
\wt M^{T}_{\rm ST_R}\otimes
\wt M^{T}_{\rm ST_L})$, {\bbf is at the origin of its spin defined by the two possible senses of rotation of its bisections.}}
\vskip 11pt

\begin{proof}
\Bi
\item Each \lr section (which is a closed string)
$\phi ^{T_p}_{\mu _L}$
\resp{$\phi ^{T_p}_{\mu _R}$} of
$\wt M^{T_p}_{\rm ST_L}$
\resp{$\wt M^{T_p}_{\rm ST_R}$}
has two possible senses of rotation due to the differential operator
$T_{L;{\rm ST}}$
\resp{$T_{R;{\rm ST}}$} acting on 
$\phi  _{\mu _L}$
\resp{$\phi  _{\mu _R}$} according to:
\[ \phi ^{T_p}_{\mu _L}=T_{L;{\rm ST}}(\phi _{\mu _L})
\rresp{\phi ^{T_p}_{\mu _R}=T_{R;{\rm ST}}(\phi _{\mu _R})}.\]

\item Similarly, every bisection
$(\phi ^{T_p}_{\mu _R}\otimes\phi ^{T_p}_{\mu _L})$
of $(\wt M^{T_p}_{\rm ST_R}\otimes
\wt M^{T_p}_{\rm ST_L})$ has two possible senses of rotation in such a way that every right section has an opposite sense of rotation with respect to its symmetric left equivalent.

\item Each bisection 
$(\phi ^{T_p}_{\mu _R}\otimes\phi ^{T_p}_{\mu _L})$
behaves dynamically like a harmonic oscillator as proved in \cite{Pie1} and was interpreted as a bound bisemiphoton at $\mu $ biquanta.

\item The two different senses of rotation of all the bisections of the time bosonic string field
$(\wt M^{T_p}_{\rm ST_R}\otimes
\wt M^{T_p}_{\rm ST_L})$ must be related to its spin.
\qedhere
\Ei
\end{proof}
\vskip 11pt

\subsection{Corollary}

{\em {\bbf The dynamics of the string field
$(\wt M^{T}_{\rm ST_R}\otimes
\wt M^{T}_{\rm ST_L})$, given by its operator-valued string field
$(\wt M^{T_p}_{\rm ST_R}\otimes
\wt M^{T_p}_{\rm ST_L})$, reflects its evolving nature.}}
\vskip 11pt

\begin{proof}
The dynamics of the string field
$(\wt M^{T_p}_{\rm ST_R}\otimes
\wt M^{T_p}_{\rm ST_L})$ is characterized by:
\Bena
\item quantization rules consisting in adding or removing a set of biquanta on its bisections
$(\phi ^{T_p}_{\mu _R}\otimes
\phi ^{T_p}_{\mu _L})$ as developed in proposition 2.6;

\item the rotation (in opposite senses) of its bisections which leads to an exchange of left and right quanta inside them, which is at the origin of the magnetic subfield associated with this string field
$(\wt M^{T_p}_{\rm ST_R}\otimes
\wt M^{T_p}_{\rm ST_L})$.  Indeed, this string field is defined over the algebraic bilinear semigroup
$\GL_2(F_{\o v}\times F_v)$ which allows the following splitting:
\[
(\wt M^{T_p}_{\rm ST_R}\otimes
\wt M^{T_p}_{\rm ST_L})\overset{\sim}{\To}
(\wt M^{T_p}_{\rm ST_R}\otimes_{D}
\wt M^{T_p}_{\rm ST_L})\oplus
(\wt M^{T_p}_{\rm ST_R}\otimes_m
\wt M^{T_p}_{\rm ST_L})\]
generating the magnetic subfield
$(\wt M^{T_p}_{\rm ST_R}\otimes_m
\wt M^{T_p}_{\rm ST_L})$ as indicated in propositions 2.3 and 2.4 and in \cite{Pie5} on the basis of Green's propagators in the Feynman's diagram framework.
\Ee

The existence of quantization rules (and of a magnetic subfield) is thus at the origin of the evolution of the (operator-valued) string field
$(\wt M^{T_p}_{\rm ST_R}\otimes
\wt M^{T_p}_{\rm ST_L})$ since its structure, given by the number of its biquanta, changes with respect to the clock time.
\end{proof}
\vskip 11pt

\subsection{Endomorphisms on semifields}

Two kinds of smooth endomorphisms $E$ and $E_t$ can be applied on the \lr semifield
$\wt M^{T_p}_{\rm ST_L}$
\resp{$\wt M^{T_p}_{\rm ST_R}$} \cite{Pie1}:

\Bean
\item {\bbf The first endomorphism:}
\begin{align*}
E_L: \qquad \wt M^{T_p}_{\rm ST_L}&\To
\wt M^{T_p(r)}_{\rm ST_L}\oplus
\wt M^{T_p(i)}_{\rm ST_L}\\[8pt]
\rresp{
E_R: \qquad \wt M^{T_p}_{\rm ST_R}&\To
\wt M^{T_p(r)}_{\rm ST_R}\oplus
\wt M^{T_p(i)}_{\rm ST_R}}\end{align*}
where:\Bi
\item $\wt M^{T_p(r)}_{\rm ST_L}$
\resp{$\wt M^{T_p(r)}_{\rm ST_R}$} is the residue time semifield of which algebraic semigroup, on which it is defined, is submitted to Galois antiautomorphisms;

\item $\wt M^{T_p(i)}_{\rm ST_L}$
\resp{$\wt M^{T_p(i)}_{\rm ST_R}$}
is the complementary time semifield of which algebraic semigroup, on which it is defined, is submitted to Galois automorphisms;
\Ei

{\bbf is such that  $\wt M^{T_p(r)}_{\rm ST_L}$
\resp{$\wt M^{T_p(r)}_{\rm ST_R}$} and
 $\wt M^{T_p(i)}_{\rm ST_L}$
\resp{$\wt M^{T_p(i)}_{\rm ST_R}$} are assumed to be not connected.}

Then,  $\wt M^{T_p(i)}_{\rm ST_L}$
\resp{$\wt M^{T_p(i)}_{\rm ST_R}$} is mapped through the origin into a three-dimensional complementary space semifield
 $\wt M^{S_p}_{\rm ST_L}$
\resp{$\wt M^{S_p}_{\rm ST_R}$} according to:
\[ \gamma _{t\to r} : \quad \wt M^{T_p(i)}_{\rm ST_L}\To \wt M^{S_p }_{\rm ST_L}
\rresp{\gamma _{t\to r} : \quad \wt M^{T_p(i)}_{\rm ST_R}\To \wt M^{S_p }_{\rm ST_R}}.\]
So, we have that:
\begin{align*}
 \gamma _{t\to r} \circ E_L: \qquad \wt M^{T_p}_{\rm ST_L}&\To \wt M^{T_p(r) }_{\rm ST_L}\oplus \wt M^{S_p }_{\rm ST_L}\\[8pt]
 \rresp{\gamma _{t\to r} \circ E_R: \qquad \wt M^{T_p}_{\rm ST_R}&\To \wt M^{T_p(r) }_{\rm ST_R}\oplus \wt M^{S_p }_{\rm ST_R}}\end{align*}
 in such a way that
 $(\wt M^{S_p }_{\rm ST_R}\otimes \wt M^{S_p }_{\rm ST_L})$ is a ``space'' field of which bielements are rotating closed bistrings localized in a (bisemi)space orthogonal to the closed bistrings of the time ``residue'' field
$( \wt M^{T_p(r) }_{\rm ST_R}\otimes\wt M^{T_p(r) }_{\rm ST_L})$.

\item {\bbf The second endomorphism $E_t$ decomposes the \lr semifield
$\wt M^{T_p }_{\rm ST_L}$
\resp{$\wt M^{T_p }_{\rm ST_R}$} into the residue time semifield
$\wt M^{T_p(r) }_{\rm ST_L}$
\resp{$\wt M^{T_p(r) }_{\rm ST_R}$} and into the complementary time semifield
$\wt M^{T_p(i) }_{\rm ST_L}$
\resp{$\wt M^{T_p(i) }_{\rm ST_R}$} which are assumed to be connected.}

This endomorphism $E_t$ was interpreted in \cite{Pie1} as being at the origin of the generation of the time internal semifields of the (semi)quarks from a ``core'' time semifield
$\wt M^{T_p}_{\rm ST_L}$ \resp{$\wt M^{T_p}_{\rm ST_R}$}, then rewritten according to
$\wt M^{{\rm (Bar)};T}_{\rm ST_L}$
\resp{$\wt M^{{\rm (Bar)};T}_{\rm ST_R}$}.

Indeed, $E_t$, composed with a $\gamma _{t\to\oplus t_i}$ morphism sending the complementary time semifield into a $3D$  complementary semispace which is of time type due to its connectedness, transforms
$\wt M^{{\rm (Bar)};T}_{\rm ST_L}$
\resp{$\wt M^{{\rm (Bar)};T}_{\rm ST_R}$} into:
\begin{align*}
\gamma _{t\to\oplus t_i}\circ E_{t_L}: \qquad
\wt M^{{\rm (Bar)};T}_{\rm ST_L}
&\To
\wt M^{{\rm (Bar)};T(r)}_{\rm ST_L}\bigoplus_{i=1}^3
\wt M^{q_i;T}_{\rm ST_L}\\[8pt]
\rresp{
\gamma _{t\to\oplus t_i}\circ E_{t_R}: \qquad
\wt M^{{\rm (Bar)};T}_{\rm ST_R}
&\To
\wt M^{{\rm (Bar)};T(r)}_{\rm ST_R}\bigoplus_{i=1}^3
\wt M^{q_i;T}_{\rm ST_R}}
\end{align*}
where:
\Bi
\item $\wt M^{{\rm (Bar)};T(r)}_{\rm ST_L}$
\resp{$\wt M^{{\rm (Bar)};T(r)}_{\rm ST_R}$} is the core residue time semifield of a \lr semibaryon;

\item $\wt M^{q_i;T}_{\rm ST_L}$
\resp{$\wt M^{q_i;T}_{\rm ST_R}$} is the internal time string semifield of a \lr semiquark.
\Ei

The time string semifield of the three \lr semiquarks are then connected to the (residue) core time semifield of the considered \lr semibaryon, which explains the confinement of these semiquarks into a ``bag'' leading to an algebraic interpretation of the constant of the strong interaction (at this space-time level) in terms of algebraic Hecke characters as developed in \cite{Pie1}.

Note that the space semifield
$\wt M^{q_i;S}_{\rm ST_L}$
\resp{$\wt M^{q_i;S}_{\rm ST_R}$} of the three semiquarks $q_i$, $1\le i\le 3$, at this internal vacuum level, can be generated from their time string semifields
$\wt M^{q_i;T}_{\rm ST_L}$
\resp{$\wt M^{q_i;T}_{\rm ST_R}$} by
$(\gamma _{t_{i_L}\to r_{i_L}}\circ E_{i_L})$
\resp{$(\gamma _{t_{i_R}\to r_{i_R}}\circ E_{i_R})$}
morphisms as explained in a) with $E_{i_L}$ \resp{$E_{i_R}$} a smooth endomorphism of the first kind.

So, the {\bbf space-time internal vacuum of a \lr semibaryon can be developed according to:}
\begin{align*}
\wt M^{{\rm (Bar)};T-S}_{\rm ST_L}&=
\wt M^{{\rm (Bar)};T}_{\rm ST_L}\bigoplus_{i=1}^3
\wt M^{q_i;T-S}_{\rm ST_L}\\[8pt]
\rresp{\wt M^{{\rm (Bar)};T-S}_{\rm ST_R}&=
\wt M^{{\rm (Bar)};T}_{\rm ST_R}\bigoplus_{i=1}^3
\wt M^{q_i;T-S}_{\rm ST_R}}.
\end{align*}
\Ee
\vskip 11pt

All that can be summarized in the following proposition.
\vskip 11pt

\subsection{Proposition}
{\em
\Bean
\item {\bbf The space internal vacuum string semifield of a simple semifermion, i.e. a semilepton or a semiquark, is generated from its time string semifield under a $(\gamma _{t\to r}\circ E)$ morphism} where $E$ is a smooth endomorphim of the first kind.

\item {\bbf The core time internal vacuum string semifield of a composite semifermion, i.e. a semibaryon, can generate the time string semifields of the three semiquarks} at this ST level under $(\gamma _{t\to \oplus t_i}\circ E_t)$ morphisms, where $E_t$ is a smooth endomorphism of the second kind, the space string semifields of the semiquarks are generated from their time string semifields by
$(\gamma _{t_i\to r_i}\circ E_i)$ morphisms.

But, the generation of the time string semifields of the three semiquarks from a core time string semifield (of a semibaryon) with respect to the generation of a space string semifield of a semilepton of the same family is only possible if the number of sections $n_{t_{b_j}}$ of the core time string semifield of the semibaryon is superior to the number of sections $n_{t_{\ell_j}}$ of the time string semifield of the envisaged semilepton.
\Ee}
\vskip 11pt

\begin{proof} It is evident that 
$n_{t_{b_j}} \gg n_{t_{\ell_j}}$, where $1\le j\le 3$ refers to the three possible families, since, by construction, the generation of time and space time string semifields of semiquarks requires much more sections, i.e. quanta, than the generation of the space string semifield of the semilepton of the same family.
\end{proof}
\vskip 11pt

\subsection{Operator valued space-time ``space'' semifields}

\Bi
\item As in proposition 2.9, the operator valued string semifield
$\wt M^{S_p}_{{\rm ST}_L}$
\resp{$\wt M^{S_p}_{{\rm ST}_R}$} of a semilepton is given by the total semispace of the tangent bundle
$(\wt M^{S_p}_{{\rm ST}_L},S_{L;{\rm ST}},\wt M^{S}_{{\rm ST}_L})$ 
\resp{$(\wt M^{S_p}_{{\rm ST}_R},S_{R;{\rm ST}},\wt M^{S}_{{\rm ST}_R})$}
where:
\Bi
\item $S_{L;\rm ST}$ is the differential operator
\[ S_{L;\rm ST}=i\ \dfrac{\hbar_{\rm ST}}{c_{t\to r;\rm ST}}
\L\{ s_{x_L}\ \dfrac\partial{\partial x},s_{y_L}\ \dfrac\partial{\partial y},s_{z_L}\ \dfrac\partial{\partial z}\R\}\]
with $\vec{s_L}=\L\{s_{x_L},s_{y_L},s_{z_L}\R\}$ the vector responsible for the spin direction in the above directional derivative,

and $S_{R;\rm ST}$ is similarly the differential operator
\[ S_{R;\rm ST}=-i\ \dfrac{\hbar_{\rm ST}}{c_{t\to r;\rm ST}}
\L\{ s_{x_L}\ \dfrac\partial{\partial x},s_{y_L}\ \dfrac\partial{\partial y},s_{z_L}\ \dfrac\partial{\partial z}\R\}\;.\]
\Ei

\item Referring to \cite{Pie1} and \cite{Pie5}, the internal vacuum operator valued string semifields of a \lr semybaryon are given by:
\begin{align*}
\wt M^{({\rm Bar});T_p-S_p}_{{\rm ST}_L}
&=
\wt M^{({\rm Bar});T_p}_{{\rm ST}_L}\bigoplus_{i=1}^3
\wt M^{q_i;T_p-S_p}_{{\rm ST}_L}\\[8pt]
\rresp{
\wt M^{({\rm Bar});T_p-S_p}_{{\rm ST}_R}
&=
\wt M^{({\rm Bar});T_p}_{{\rm ST}_R}\bigoplus_{i=1}^3
\wt M^{q_i;T_p-S_p}_{{\rm ST}_R}}\;.\end{align*}
They are submitted to the action of the differential operator (left case):
\begin{multline*}
{\rm TS}^{({\rm Bar})}_{{\rm ST}_L}
: \L\{ 
i\hbar_{\rm ST}\ s_{0_L}\ \dfrac\partial{\partial t_{c_0}},
\L\{
 i\hbar_{\rm ST}\ G^{-1}(\rho  )\ s^{(1)}_{0_L}\ 
\dfrac\partial{\partial t_{0}},
i\ \dfrac{\hbar_{\rm ST}}{c_{t\to r}}\ G^{-1}(\rho )\ s^{(1)}_{x_L}\ \dfrac\partial{\partial x},
\R.
\R.\\[8pt]
\L.
\dots,
i\ \dfrac{\hbar}{c_{t\to r}}\ G^{-1}(\rho )\ s^{(1)}_{z_L}\ \dfrac\partial{\partial z}\R\},\biggl\{\dots\biggr\},
\biggl\{\dots,
\L.\L.
i\ \dfrac{\hbar}{c_{t\to r}}\ G^{-1}(\rho )\ s^{(3)}_{z_L}\ \dfrac\partial{\partial z}\R\}\R\}
\end{multline*}
where:
\Bi
\item $i\hbar_{\rm ST}\ s_{0_L}\dfrac\partial{\partial t_{c_0}}$ is the differential operator acting on the core time string semifield
$\wt M^{({\rm Bar});T}_{{\rm ST}_L}$ of the left semibaryon;

\item the three internal parenthesis refer to the differential operators acting on the space-time string semifields of the three semiquarks at this ``~ST~'' internal vacuum level;

\item $G(\rho )$ is the strong coupling constant defined in the frame of AQT \cite{Pie1}.
\Ei
\Ei
\vskip 11pt

\subsection{Proposition}

{\em {\bbf The operator valued string fields of the internal vacuum ``~ ST~'' of a simple bisemifermion, i.e. a bisemilepton (or a bisemiquark), are given by:}
\begin{multline*}
(\wt M^{T_p-S_p}_{{\rm ST}_R}\otimes
\wt M^{T_p-S_p}_{{\rm ST}_L})
= 
\L[
(\wt M^{T_p}_{{\rm ST}_R}\otimes_D\wt M^{T_p}_{{\rm ST}_L})\oplus
(\wt M^{T_p}_{{\rm ST}_R}\otimes_m\wt M^{T_p}_{{\rm ST}_L})\R]\\[8pt]
\qquad \qquad \oplus
\L[
(\wt M^{S_p}_{{\rm ST}_R}\otimes_D\wt M^{T_p}_{{\rm ST}_L})
\oplus
(\wt M^{S_p}_{{\rm ST}_R}\otimes_m\wt M^{T_p}_{{\rm ST}_L})\R]\\[8pt]
\oplus
\L[ (\wt M^{(T_p)-S_p}_{{\rm ST}_R}\otimes_e\wt M^{(S_p)-T_p}_{{\rm ST}_L})\R]
\end{multline*}
where:
\Bi
\item
$\L[
(\wt M^{T_p}_{{\rm ST}_R}\otimes_D\wt M^{T_p}_{{\rm ST}_L})\oplus
(\wt M^{T_p}_{{\rm ST}_R}\otimes_m\wt M^{T_p}_{{\rm ST}_L})\R]$
{\bbf is the time string field decomposed into a diagonal time string subfield and into a possible magnetic time (string) subfield } responsible for the exchange of magnetic quanta inside time bistrings.

Remark that describing the exchange of magnetic quanta as a (magnetic) string field is only a picture of the reality {\em \cite{F-L-S}}.  It is explicitly described in {\em \cite{Pie5};}

\item $\L[
(\wt M^{S_p}_{{\rm ST}_R}\otimes_D\wt M^{S_p}_{{\rm ST}_L})\oplus
(\wt M^{S_p}_{{\rm ST}_R}\otimes_m\wt M^{S_p}_{{\rm ST}_L})\R]$ {\bbf is the space string field decomposed similarly into diagonal and magnetic space string subfields;}

\item $\L[ (\wt M^{(T_p)-S_p}_{{\rm ST}_R}\otimes_e\wt M^{(S_p)-T_p}_{{\rm ST}_L})\R]$ {\bbf is the electric string field responsible for the electric charge} of the bisemilepton of this ST level by the exchange of electric quanta between the left semifield of time (resp. space) and the right semifield of space (resp. time).
\Ei
}
\vskip 11pt

\begin{proof}
The existence of magnetic subfields of time and space type results from the fact that the time and space string fields 
$(\wt M^{T_p}_{{\rm ST}_R}\otimes\wt M^{T_p}_{{\rm ST}_L})$ and
$(\wt M^{S_p}_{{\rm ST}_R}\otimes\wt M^{S_p}_{{\rm ST}_L})$ are defined on algebraic bilinear semigroups of which representation spaces split into  diagonal and  magnetic tensor products according to proposition 2.3.
\end{proof}
\vskip 11pt

\subsection{Proposition}

{\em {\bbf The operator valued string fields of the internal vacuum ``~ST~'' of a composite bisemifermion, i.e. a bisemibaryon, are given by:}
\begin{multline*}
(\wt M^{({\rm Bar});T_p-S_p}_{{\rm ST}_R}
\otimes
\wt M^{({\rm Bar});T_p-S_p}_{{\rm ST}_L})
=
(\wt M^{({\rm Bar});T_p(r)}_{{\rm ST}_R}
\otimes
\wt M^{({\rm Bar});T_p(r)}_{{\rm ST}_L})\\[8pt]
\begin{aligned}\bigoplus_{i=1}^3
(\wt M^{q_i;T_p-S_p}_{{\rm ST}_R}\otimes
\wt M^{q_i;T_p-S_p}_{{\rm ST}_L})
\bigoplus_{\begin{subarray}{c}
i=j=1\\ i\neq j \end{subarray}}^3
(\wt M^{q_i;T_p-S_p}_{{\rm ST}_R}\otimes
\wt M^{q_j;T_p-S_p}_{{\rm ST}_L})\\[8pt]
\bigoplus_{i=1}^3
(\wt M^{{\rm (Bar)};T_p(r)}_{{\rm ST}_R}\otimes
\wt M^{q_i;T_p-S_p}_{{\rm ST}_L})
\bigoplus_{i=1}^3
(\wt M^{q_i;T_p-S_p}_{{\rm ST}_R}\otimes
\wt M^{{\rm (Bar)};T_p(r)}_{{\rm ST}_L})
\end{aligned}
\end{multline*}
where:
\Bean
\item $(\wt M^{({\rm Bar});T_p(r)}_{{\rm ST}_R}
\otimes
\wt M^{({\rm Bar});T_p(r)}_{{\rm ST}_L})$ is the {\bbf core central time string field of the bisemibaryon};

\item $(\wt M^{q_i;T_p-S_p}_{{\rm ST}_R}\otimes
\wt M^{q_i;T_p-S_p}_{{\rm ST}_L})$ is the {\bbf internal vacuum string field of the $i$-th bisemiquark} decomposing as it was done in proposition 2.14, into diagonal space and time string fields, into magnetic space and time (string) fields and into an electric (string) field responsible for an electric charge at this level having an absolute value 
$\L|\tfrac13\R|\ e$
or $\L|\tfrac23\R|\ e$;

\item $(\wt M^{q_i;T_p-S_p}_{{\rm ST}_R}\otimes
\wt M^{q_j;T_p-S_p}_{{\rm ST}_L})$ are {\bbf mixed ``string'' fields of interaction between the $i$-th right semiquark and the $j$-th left semiquark.  They decompose similarly into:}
\Bi
\item {\bbf diagonal space and time ``string'' fields which are gravitational}, i.e. they exchange gravitational quanta as it was seen in {\em \cite{Pie1};}

\item {\bbf magnetic space and time fields of interaction};

\item {\bbf an electric ``string'' field of interaction}.
\Ei

\item $(\wt M^{{\rm (Bar)};T_p(r)}_{{\rm ST}_R}\otimes
\wt M^{q_i;T_p-S_p}_{{\rm ST}_L})$ (and, inversely,
$(\wt M^{q_i;T_p-S_p}_{{\rm ST}_R}\otimes
\wt M^{{\rm (Bar)};T_p(r)}_{{\rm ST}_L})$) is the {\bbf ``strong'' mixed (string) field of interaction} between the right core central time string semifield of the right semibaryon and the space-time string semifield of the $i$-th left semiquark.

It decomposes according to:
\[
(\wt M^{{\rm (Bar)};T_p(r)}_{{\rm ST}_R}\otimes
\wt M^{q_i;T_p-S_p}_{{\rm ST}_L})
=
(\wt M^{{\rm (Bar)};T_p(r)}_{{\rm ST}_R}\otimes
\wt M^{q_i;T_p}_{{\rm ST}_L})
\oplus
(\wt M^{{\rm (Bar)};T_p(r)}_{{\rm ST}_R}\otimes_e
\wt M^{q_i;S_p}_{{\rm ST}_L})
\]
where $(\wt M^{{\rm (Bar)};T_p(r)}_{{\rm ST}_R}\otimes
\wt M^{q_i;T_p}_{{\rm ST}_L})$ and
$(\wt M^{{\rm (Bar)};T_p(r)}_{{\rm ST}_R}\otimes_e
\wt M^{q_i;S_p}_{{\rm ST}_L})$
are respectively {\bbf mixed strong gravitational and electric (string) fields of interaction.}
\Ee
}
\vskip 11pt

\begin{proof}
These gravitational and electric (string) fields of interaction
$(\wt M^{{\rm (Bar)};T_p(r)}_{{\rm ST}_R}\otimes
\wt M^{q_i;T_p}_{{\rm ST}_L})$ and
$(\wt M^{{\rm (Bar)};T_p(r)}_{{\rm ST}_R}\otimes_e
\wt M^{q_i;S_p}_{{\rm ST}_L})$
are of strong nature because they contribute to the confinement of the (bi)semiquarks by their binding to the time core central structure
$(\wt M^{{\rm (Bar)};T_p(r)}_{{\rm ST}_R}\otimes
\wt M^{{\rm (Bar)};T_p(r)}_{{\rm ST}_L})$ of the bisemibaryon.
\end{proof}
\vskip 11pt

\subsection{Proposition}

{\em {\bbf The (string) strong fields of interaction inside a perturbed bisemibaryon can condense into massive mesons.}}
\vskip 11pt

\begin{proof}
\Bi
\item These mixed (string) strong fields at the ``~ST~'' level are
$(\wt M^{{\rm (Bar)};T_p(r)}_{{\rm ST}_R}\otimes
\wt M^{q_i;T_p-S_p}_{{\rm ST}_L})$ and
$(\wt M^{q_j;T_p-S_p}_{{\rm ST}_R} \otimes
\wt M^{{\rm (Bar)};T_p(r)}_{{\rm ST}_L})$
referring to proposition 2.15 d).

\item Assume that the set of exchange \rl time strings of
$(\wt M^{{\rm (Bar)};T_p(r)}_{{\rm ST}_R})$
\resp{$
(\wt M^{{\rm (Bar)};T_p(r)}_{{\rm ST}_L})$}, composed of \rl exchanged quanta, generates their corresponding space strings by
$(\gamma _{t\to r}\circ E_R)$
\resp{$(\gamma _{t\to r}\circ E_L)$}
morphisms (see section 2.11).

\item So, we get a set of ``$\beta $'' exchanged right (resp. ``$\gamma $'' exchanged left) space-time strings of
$(\wt M^{{\rm (Bar)};T_p-S_p}_{{\rm ST}_R})$
\resp{$
(\wt M^{{\rm (Bar)};T_p-S_p}_{{\rm ST}_L})$} joining a corresponding set of 
``$\beta $'' exchanged left (resp. ``$\gamma $'' exchanged right) space-time strings of
$\wt M^{q_i;T_p-S_p}_{{\rm ST}_R}$
\resp{$
\wt M^{q_j;T_p-S_p}_{{\rm ST}_L}$} leading to the sum of fields:
\[
(\wt M^{{\rm (Bar)};T_p-S_p}_{{\rm ST}_R}\{\beta \}
\otimes
\wt M^{q_i;T_p-S_p}_{{\rm ST}_L}\{\beta \})
\oplus
(\wt M^{q_j;T_p-S_p}_{{\rm ST}_R}\{\gamma  \}
\otimes
\wt M^{{\rm (Bar)};T_p-S_p}_{{\rm ST}_L}\{\gamma  \})\;.\]

\item The condensation, i.e. the compactification of the interaction  ``string'' field
$(\wt M^{{\rm (Bar)};T_p-S_p}_{{\rm ST}_R}\{\beta \}\linebreak
\otimes
\wt M^{q_i;T_p-S_p}_{{\rm ST}_L}\{\beta \})$
\resp{$(\wt M^{q_j;T_p-S_p}_{{\rm ST}_R}\{\gamma  \}
\otimes
\wt M^{{\rm (Bar)};T_p-S_p}_{{\rm ST}_L}\{\gamma  \})$}
decomposing into diagonal, magnetic and electric string fields, has the structure of a bisemiquark
$q_i$ \resp{$\o{q_j}$} in such a way that their direct sum
$q_i\oplus \o{q_j}$, written in condensed form
$\o{q_j}\ q_i$, is a meson having the (bisemi) quark structure $\o{q_j}\ q_i$ at the ``~ST~'' level of its internal vacuum structure.

This can occur, for example, after the collision of a ``high energy'' proton with another proton or in the decay of a heavy baryon into a lighter one and into a meson (i.e. $\Lambda \to p+\pi ^-$).\qedhere
\Ei
\end{proof}
\vskip 11pt

\subsection{Corollary}

{\em The generation of mesons from a perturbed bisemibaryon allows the bisemiquarks to have different quantum states.}
\vskip 11pt

\begin{proof} This is evident by construction according to proposition 2.16, preventing three identical bisemiquarks from violating the Pauli principle.  Indeed, the quark composition ($sss$) of the baryon $\Omega ^-$ led to the introduction of the color in quantum chromodynamics.
\end{proof}
\vskip 11pt

\subsection{Middle ground and mass string fields}

According to \cite{Pie1} and \cite{Pie4}, the time and space string fields of the internal vacuum structure ``~ST~'' of a bisemilepton and of a bisemibaryon are submitted to strong fluctuations generating on their sections degenerate singularities which are able to produce by versal deformations and blowups of these {\bbf two new embedded covering string fields at one sheet over the ST string fields.}

Let $ (\wt M^{T_p-S_p}_{{\rm ST}_R}\otimes
\wt M^{T_p -S_p}_{{\rm ST}_L})$ denote the operator valued space and time string fields of the internal vacuum ``~ST~''  of a bisemilepton.

Then, {\bbf the middle ground and mass string fields}
$ (\wt M^{T_p-S_p}_{{\rm MG}_R}\otimes
\wt M^{T_p -S_p}_{{\rm MG}_L})$
and
$ (\wt M^{T_p-S_p}_{{\rm M}_R}\otimes
\wt M^{T_p -S_p}_{{\rm M}_L})$
{\bbf are one-sheet covering string fields above
$ (\wt M^{T_p-S_p}_{{\rm ST}_R}\otimes
\wt M^{T_p -S_p}_{{\rm ST}_L})$} and are embedded according to:
\[
(\wt M^{T_p-S_p}_{{\rm ST}_R}\otimes
\wt M^{T_p -S_p}_{{\rm ST}_L})
\subset
(\wt M^{T_p-S_p}_{{\rm MG}_R}\otimes
\wt M^{T_p -S_p}_{{\rm MG}_L})
\subset
(\wt M^{T_p-S_p}_{{\rm M}_R}\otimes
\wt M^{T_p -S_p}_{{\rm M}_L})\;.\]
\vskip 11pt

\subsection{Proposition}

{\em {\bbf The generation of the mass string field
$ (\wt M^{T_p-S_p}_{{\rm M}_R}\otimes
\wt M^{T_p -S_p}_{{\rm M}_L})$
\resp{$ (\wt M^{{\rm (Bar)};T_p-S_p}_{{\rm M}_R}\otimes
\wt M^{{\rm (Bar)};T_p -S_p}_{{\rm M}_L})$}
from the internal vacuum}
\begin{align*}
(\wt M^{T_p-S_p}_{{\rm ST}_R}\otimes
\wt M^{T_p -S_p}_{{\rm ST}_L})
&\oplus
(\wt M^{T_p-S_p}_{{\rm MG}_R}\otimes
\wt M^{T_p -S_p}_{{\rm MG}_L})\\[8pt]
\rresp{
(\wt M^{{\rm (Bar)};T_p-S_p}_{{\rm ST}_R}\otimes
\wt M^{{\rm (Bar)};T_p -S_p}_{{\rm ST}_L})
&\oplus
(\wt M^{{\rm (Bar)};T_p-S_p}_{{\rm MG}_R}\otimes
\wt M^{{\rm (Bar)};T_p -S_p}_{{\rm MG}_L})}
\end{align*}
{\bbf of a bisemilepton \resp{a bisemibaryon} is equivalent to the Higgs mechanism} {\em \cite{B-E}, \cite{Hig},}
responsible for the generation of masses in non-abelian Gauge theories {\em \cite{K-L}.}
}
\vskip 11pt

\begin{proof} In fact, the Higgs mechanism corresponds in AQT to the generation of the mass string field
$(\wt M^{T_p}_{{\rm M}_R}\otimes
\wt M^{T_p }_{{\rm M}_L})$
\resp{$(\wt M^{{\rm (Bar)};T_p}_{{\rm M}_R}\otimes
\wt M^{{\rm (Bar)};T_p -S_p}_{{\rm M}_L})$}
of ``time'' type from its corresponding middle-ground equivalent
$(\wt M^{T_p}_{{\rm MG}_R}\otimes
\wt M^{T_p}_{{\rm MG}_L})$
\resp{$(\wt M^{{\rm (Bar)};T_p}_{{\rm MG}_R}\otimes
\wt M^{{\rm (Bar)};T_p }_{{\rm MG}_L})$}
by versal deformation ${\rm Vd}(2)$, and spreading-out isomorphism ${\rm SOT}(2)$ according to \cite{Pie1}:
\begin{align*}
{\rm SOT}(2)\circ {\rm Vd}(2)&: \qquad 
&(\wt M^{T_p }_{{\rm MG}_R}\otimes
\wt M^{T_p }_{{\rm MG}_L})
&\To
(\wt M^{T_p }_{{\rm M}_R}\otimes
\wt M^{T_p  }_{{\rm M}_L})\\[8pt]
\rresp{
{\rm SOT}(2)\circ {\rm Vd}(2)&: \qquad 
&(\wt M^{{\rm (Bar)};T_p }_{{\rm MG}_R}\otimes
\wt M^{{\rm (Bar)};T_p }_{{\rm MG}_L})
&\To
(\wt M^{{\rm (Bar)};T_p }_{{\rm M}_R}\otimes
\wt M^{{\rm (Bar)};T_p  }_{{\rm M}_L})}.
\end{align*}
The corresponding mass string field of ``space'' type is generated from the mass string field of ``time'' type by a
$\gamma ^M_{t\to r}\circ E$ morphism as described in proposition 2.12.
\end{proof}
\vskip 11pt

\subsection{Generation of bisemifermions of the 2nd and 3rd families}

\Bi
\item It will be seen how elementary bisemifermions of the second and third families can be generated from elementary bisemifermions of the first family, i.e. $e^-$, $u^+$ and $d^-$.

\item Referring to \cite{Pie5}, let
$(^A\wt M^{T_p-S_p}_{{\rm ST}_R-{\rm MG}_R-{\rm M}_R}\otimes
{}^A\wt M^{T_p-S_p}_{{\rm ST}_L-{\rm MG}_L-{\rm M}_L})$
denote the internal ``~ST~'', ``~MG~'' and ``~M~'' string fields of an elementary bisemifermion ``$A$'' of the first family.

{\bbf Under some external strong field, all the bisections} of the internal string fields ``~ST~'', ``~MG~'' and ``~M~'' of this bisemifermion ``$A$'' {\bbf are assumed to be endowed with singularities of corank $1$ and multiplicity $<3$ in such a way that the versal deformation ${\rm Vd}(B)$ and spreading-out morphisms ${\rm SOT}(B)$ of these generate covering string fields
``~ST~'', ``~MG~'' and ``~M~''
at multiple sheets above the original string fields of the bisemifermion $A$} according to:
\begin{multline*}
{\rm SOT}(B) \circ {\rm Vd}(B): \quad
(^A\wt M^{T_p-S_p}_{{\rm ST}_R-{\rm MG}_R-{\rm M}_R}\otimes
{}^A\wt M^{T_p-S_p}_{{\rm ST}_L-{\rm MG}_L-{\rm M}_L})\\[8pt]
\qquad \qquad \To
(^A\wt M^{T_p-S_p}_{{\rm ST}_R-{\rm MG}_R-{\rm M}_R}\otimes
{}^A\wt M^{T_p-S_p}_{{\rm ST}_L-{\rm MG}_L-{\rm M}_L})
\\[8pt]
\oplus
(^{A'}\wt M^{T_p-S_p}_{{\rm ST}_R-{\rm MG}_R-{\rm M}_R}\otimes
^{A'}\wt M^{T_p-S_p}_{{\rm ST}_L-{\rm MG}_L-{\rm M}_L})
\end{multline*}
in such a way that
\begin{multline*}
(^B\wt M^{T_p-S_p}_{{\rm ST}_R-{\rm MG}_R-{\rm M}_R}\otimes
{}^B\wt M^{T_p-S_p}_{{\rm ST}_L-{\rm MG}_L-{\rm M}_L})
\\[8pt]
=
(^A\wt M^{T_p-S_p}_{{\rm ST}_R-{\rm MG}_R-{\rm M}_R}\otimes
{}^A\wt M^{T_p-S_p}_{{\rm ST}_L-{\rm MG}_L-{\rm M}_L})
\\[8pt]
\oplus
(^{A'}\wt M^{T_p-S_p}_{{\rm ST}_R-{\rm MG}_R-{\rm M}_R}\otimes
{}^{A'}\wt M^{T_p-S_p}_{{\rm ST}_L-{\rm MG}_L-{\rm M}_L})
\end{multline*}
where:
\Bi
\item $(^B\wt M^{T_p-S_p}_{{\rm ST}_R-{\rm MG}_R-{\rm M}_R}\otimes
{}^B\wt M^{T_p-S_p}_{{\rm ST}_L-{\rm MG}_L-{\rm M}_L})$
are the string fields ``~ST~'', ``~MG~'' and ``~M~'' of a bisemifermion ``$B$'' of the second family;

\item $(^{A'}\wt M^{T_p-S_p}_{{\rm ST}_R-{\rm MG}_R-{\rm M}_R}\otimes
^{A'}\wt M^{T_p-S_p}_{{\rm ST}_L-{\rm MG}_L-{\rm M}_L})$
are  ``~ST~'', ``~MG~'' and ``~M~'' covering string fields at multiple sheets.
\Ei

\item If the external field is still stronger, then the bisections of the string fields
``~ST~'', ``~MG~'' and ``~M~'' of a bisemifermion $A$ will be endowed with singularities of corank $1$ and multiplicity $3$ in such a way that two covering string fields
``~ST~'', ``~MG~'' and ``~M~'' labelled ``$A'$'' and ``$A''$'', resulting from versal deformations 
${\rm Vd}(C)$ and spreading-out morphisms ${\rm SOT}(C)$, are generated above the original string fields of the bisemifermion
``$A$'' according to:
\begin{multline*}
{\rm SOT}(C) \circ {\rm Vd}(C): \quad
(^A\wt M^{T_p-S_p}_{{\rm ST}_R-{\rm MG}_R-{\rm M}_R}\otimes
{}^A\wt M^{T_p-S_p}_{{\rm ST}_L-{\rm MG}_L-{\rm M}_L})\\[8pt]
\To
(^C\wt M^{T_p-S_p}_{{\rm ST}_R-{\rm MG}_R-{\rm M}_R}\otimes
{}^C\wt M^{T_p-S_p}_{{\rm ST}_L-{\rm MG}_L-{\rm M}_L})
\end{multline*}
where
\begin{multline*}
(^C\wt M^{T_p-S_p}_{{\rm ST}_R-{\rm MG}_R-{\rm M}_R}\otimes
{}^C\wt M^{T_p-S_p}_{{\rm ST}_L-{\rm MG}_L-{\rm M}_L})
\\[8pt]
= (^A\wt M^{T_p-S_p}_{{\rm ST}_R-{\rm MG}_R-{\rm M}_R}\otimes
{}^A\wt M^{T_p-S_p}_{{\rm ST}_L-{\rm MG}_L-{\rm M}_L})
\\[8pt]
\qquad \qquad \oplus
(^{A'}\wt M^{T_p-S_p}_{{\rm ST}_R-{\rm MG}_R-{\rm M}_R}\otimes
{}^{A'}\wt M^{T_p-S_p}_{{\rm ST}_L-{\rm MG}_L-{\rm M}_L})
\\[8pt]
\oplus
(^{A''}\wt M^{T_p-S_p}_{{\rm ST}_R-{\rm MG}_R-{\rm M}_R}\otimes
{}^{A''}\wt M^{T_p-S_p}_{{\rm ST}_L-{\rm MG}_L-{\rm M}_L})
\end{multline*}
are the internal ``~ST~'', ``~MG~'' and ``~M~'' string fields of a bisemifermion ``$C$'' of the third family.
\Ei
\vskip 11pt

\section{(Strong) GEM bilinear interactions between bisemistructures of bisemiparticles}

\subsection{Internal structure and classification of bisemiparticles}

The classification of bisemiparticles is based on the two  first following lemmas.
\vskip 11pt

\subsubsection{Lemma (Bosonic fields)}

{\em \Bean
\item {\bbf The internal structure of bisemifermions is given by ``diagonal'' bosonic string fields}
$(\wt M_R\otimes_{(D)} \wt M_L)$ at the ``~ST~'', ``~MG~'' or ``~M~'' level. Each ``diagonal'' bosonic string field is composed of a family of bound ``time'' bisemibosons or ``space'' bisemibosons (i.e. ``bound'' (bisemi)photons) fitted in bistrings characterized by an increasing number of biquanta.

These bisemibosons becoming free (or ``unbound'') are then viewed as free energy at the considered level(s).

\item {\bbf The other simple bosonic fields are magnetic, electric and gravitational fields of interaction} at the ``~ST~'', ``~MG~'' and ``~M~'' levels.

\item {\bf The composite bosonic fields are massive mesons} responsible for the bilinear strong interaction.
\Ee}
\vskip 11pt

\begin{proof}
This lemma results from the developments of chapter 2. Let us only notice that:
\Bi
\item {\bbf A bisemiphoton} at the ``~ST~'', ``~MG~'' or ``~M~'' level is a ``space'' bistring of the bosonic string field
$(\wt M^{S_p}_R\otimes_{(D)} \wt M^{S_p}_L)$
at the condition that this bistring becomes free under some endomorphisms as developed in \cite{Pie1}.

It is then able to play a part in a biconnection.

\item {\bbf A magnetic simple bosonic (string) field} results from the exchange of magnetic quanta between \lr (space) strings of a \lr semifield and \rl (space) strings of a \rl semifield, as developed in \cite{Pie5}.

A set of bound magnetic biquanta is a magnetic bisemiboson.

\item {\bbf An electric simple bosonic (string) field} results from the exchange of electric quanta between \lr ``time'' \resp{``space''} strings of a \lr  ``time'' \resp{``space''} semifield and \rl ``space'' \resp{``time''} strings of a \rl  ``space'' \resp{``time''} semifield.

\item {\bbf A gravitational simple bosonic (string) field} results from the ``diagonal'' exchange of gravitational quanta between
\lr ``space'' or ``time'' strings of a  \lr semifield and 
\rl ``space'' or ``time'' strings of a \rl   semifield
of different semiparticles \cite{Pie1}.\qedhere
\Ei
\end{proof}
\vskip 11pt

\subsubsection{Lemma (Fermionic fields)}

{\em \Bean
\item {\bbf The simple bisemifermions,}
i.e. the (bisemi)leptons (and their associated neutrinos)
and the (bisemi)quarks of the three families,
{\bbf are characterized by an internal electric fields}
(at the ``~ST~'', ``~MG~'' and ``~M~'' levels)
originating from internal fields of
 ``time'' type (at these levels).

\item {\bbf The composite bisemifermions},
i.e. the (bisemi)baryons,
{\bbf are characterized by a core central field of ``time'' type}
responsible for interaction massive strong string fields essentially of electric and gravitational nature.
\Ee}
\vskip 11pt

\begin{proof}[Sketch of proof]
The fermionic character then results directly from the existence of ``time'' string fields responsible for the generation of internal electric fields between internal \lr semifields of 
``space'' \resp{``time''} type
and \rl semifields of ``time'' \resp{``space''} type.  These interacting fields then hold confined all the internal
``space'' and ``time'' fields of a bisemifermion.
\end{proof}
\vskip 11pt

\subsubsection{Lemma (Quantum statistics)}

{\em\bbf
\Bi
\item The bisemifermions obey the Fermi-Dirac statistics.
\item The bisemibosons obey the Bose-Einstein statistics.
\Ei
}
\vskip 11pt

\begin{proof}
\Bi
\item As the simple bisemifermions are held confined by means of internal electric fields responsible for their charges, they cannot congregate together at a same quantum state as it can be the case for simple ``diagonal'', gravitational, magnetic and electric bisemibosons according to lemma 3.1.

\item What is true for simple bisemifermions is evident for composite bisemifermions, i.e. bisemi\-baryons. On the other hand, the strong bisemibosons at the basis of the mixed interacting strong fields generating mesons can congregate together as developed in proposition 2.15 and thus obey the Bose-Einstein statistics.\qedhere
\Ei
\end{proof}
\vskip 11pt

\subsubsection{Internal structure of simple massive bisemifermions}

The simple massive bisemifermions, i.e. the three leptons
$e^-$, $\mu ^-$, $\tau ^-$ (and their neutrinos) and the six quarks
$u^+$, $d^-$, $s^-$, $c^+$, $b^-$, $t^+$, are characterized by the bisemistructure
$(\wt M^{T_p-S_p}_{{\rm ST}_R-{\rm MG}_R-{\rm M}_R}\otimes
\wt M^{T_p-S_p}_{{\rm ST}_L-{\rm MG}_L-{\rm M}_L})$
which decomposes according to the following string fields:
\begin{multline*}
(\wt M^{T_p-S_p}_{{\rm ST}_R-{\rm MG}_R-{\rm M}_R}\otimes
\wt M^{T_p-S_p}_{{\rm ST}_L-{\rm MG}_L-{\rm M}_L})
\\[8pt]
\equiv
(\wt M^{T_p-S_p}_{{\rm ST}_R}\oplus
\wt M^{T_p-S_p}_{{\rm MG}_R}\oplus
\wt M^{T_p-S_p}_{{\rm M}_R})
\otimes
(\wt M^{T_p-S_p}_{{\rm ST}_L}\oplus
\wt M^{T_p-S_p}_{{\rm MG}_L}\oplus
\wt M^{T_p-S_p}_{{\rm M}_L})
\\[8pt]
= \begin{aligned}[t]
(\wt M^{T_p-S_p}_{{\rm ST}_R}\otimes
\wt M^{T_p-S_p}_{{\rm ST}_L})
&\oplus
(\wt M^{T_p-S_p}_{{\rm MG}_R}\otimes
\wt M^{T_p-S_p}_{{\rm MG}_L})
\oplus
(\wt M^{T_p-S_p}_{{\rm M}_R}\otimes
\wt M^{T_p-S_p}_{{\rm M}_L})
\\[8pt]
&\oplus
(\wt M^{T_p-S_p}_{{\rm ST}_R}\otimes
\wt M^{T_p-S_p}_{{\rm MG}_L})
\oplus
(\wt M^{T_p-S_p}_{{\rm MG}_R}\otimes
\wt M^{T_p-S_p}_{{\rm ST}_L})
\\[8pt]
&\oplus
(\wt M^{T_p-S_p}_{{\rm ST}_R}\otimes
\wt M^{T_p-S_p}_{{\rm M}_L})
\oplus
(\wt M^{T_p-S_p}_{{\rm ST}_L}\otimes
\wt M^{T_p-S_p}_{{\rm M}_R})
\\[8pt]
&\oplus
(\wt M^{T_p-S_p}_{{\rm MG}_R}\otimes
\wt M^{T_p-S_p}_{{\rm M}_L})
\oplus
(\wt M^{T_p-S_p}_{{\rm M}_R}\otimes
\wt M^{T_p-S_p}_{{\rm MG}_L})
\end{aligned}
\end{multline*}
in such a way that:
\Bi
\item there is {\bf the following embedding
\[ (\wt M^{T_p-S_p}_{{\rm ST}_R}\otimes
\wt M^{T_p-S_p}_{{\rm ST}_L})
\subset
(\wt M^{T_p-S_p}_{{\rm MG}_R}\otimes
\wt M^{T_p-S_p}_{{\rm MG}_L})
\subset
(\wt M^{T_p-S_p}_{{\rm M}_R}\otimes
\wt M^{T_p-S_p}_{{\rm M}_L})\]
between the space-time, middle ground and mass string fields, each one decomposing into a diagonal, a magnetic and an electric field}
according to proposition~2.14.

\item {\bf the six other fields
\[(\wt M^{T_p-S_p}_{{\rm ST}_R}\otimes
\wt M^{T_p-S_p}_{{\rm MG}_L})\quad  \dots
\quad (\wt M^{T_p-S_p}_{{\rm M}_R}\otimes
\wt M^{T_p-S_p}_{{\rm MG}_L})\]
are interactions fields between different right and left semifields
``~ST~'', ``~MG~'' and ``~M~''.}
\Ei

They decompose into diagonal, magnetic and electric subfields responsible for the interactions between the three internal embedded shells.  In fact, only the magnetic and electric subfields between different right and left semifields
``~ST~'', ``~MG~'' and ``~M~'' are relevant since they are responsible   for the exchange of biquanta between two different shells and thus generate forces between the right and left semishells
``~ST~'' and``~MG~'',
``~ST~'' and``~M~'' and
``~MG~'' and``~M~'' as it will be seen in the following.
\vskip 11pt

\subsubsection{Internal structure of composite massive bisemifermions}

A composite massive bisemifermion, i.e. a baryon, is characterized by the bisemistructure
$(\wt M^{{\rm (Bar)};T_p-S_p}_{{\rm ST}_R-{\rm MG}_R-{\rm M}_R}\otimes
\wt M^{{\rm (Bar)};T_p-S_p}_{{\rm ST}_L-{\rm MG}_L-{\rm M}_L})$
which decomposes according to the following internal string fields:
\pagebreak

\begin{multline*}
(\wt M^{{\rm (Bar)};T_p-S_p}_{{\rm ST}_R-{\rm MG}_R-{\rm M}_R}\otimes
\wt M^{{\rm (Bar)};T_p-S_p}_{{\rm ST}_L-{\rm MG}_L-{\rm M}_L})
\\[8pt]
\begin{aligned}[t]
&= 
(\wt M^{{\rm (Bar)};T_p-S_p}_{{\rm ST}_R}\oplus
\wt M^{{\rm (Bar)};T_p-S_p}_{{\rm MG}_R}\oplus
\wt M^{{\rm (Bar)};T_p-S_p}_{{\rm M}_R})
\\[8pt]
& \qquad\otimes
(\wt M^{{\rm (Bar)};T_p-S_p}_{{\rm ST}_L}\oplus
\wt M^{{\rm (Bar)};T_p-S_p}_{{\rm MG}_L}\oplus
\wt M^{{\rm (Bar)};T_p-S_p}_{{\rm M}_L})
\\[8pt] 
&=
(\wt M^{{\rm (Bar)};T_p-S_p}_{{\rm ST}_R}\otimes
\wt M^{{\rm (Bar)};T_p-S_p}_{{\rm ST}_L})
\oplus
(\wt M^{{\rm (Bar)};T_p-S_p}_{{\rm MG}_R}\otimes
\wt M^{{\rm (Bar)};T_p-S_p}_{{\rm MG}_L})
\\[8pt]
& \qquad
\oplus
(\wt M^{{\rm (Bar)};T_p-S_p}_{{\rm M}_R}\otimes
\wt M^{{\rm (Bar)};T_p-S_p}_{{\rm M}_L})
\oplus
(\wt M^{{\rm (Bar)};T_p-S_p}_{{\rm ST}_R}\otimes
\wt M^{{\rm (Bar)};T_p-S_p}_{{\rm MG}_L})
\\[8pt]
& \qquad
\oplus
(\wt M^{{\rm (Bar)};T_p-S_p}_{{\rm MG}_R}\otimes
\wt M^{{\rm (Bar)};T_p-S_p}_{{\rm ST}_L})
\oplus
(\wt M^{{\rm (Bar)};T_p-S_p}_{{\rm ST}_R}\otimes
\wt M^{{\rm (Bar)};T_p-S_p}_{{\rm M}_L})
\\[8pt]
& \qquad
\oplus
(\wt M^{{\rm (Bar)};T_p-S_p}_{{\rm M}_R}\otimes
\wt M^{{\rm (Bar)};T_p-S_p}_{{\rm ST}_L})
\oplus
(\wt M^{{\rm (Bar)};T_p-S_p}_{{\rm MG}_R}\otimes
\wt M^{{\rm (Bar)};T_p-S_p}_{{\rm M}_L})
\\[8pt]
& \qquad
\oplus
(\wt M^{{\rm (Bar)};T_p-S_p}_{{\rm M}_R}\otimes
\wt M^{{\rm (Bar)};T_p-S_p}_{{\rm MG}_L})
\end{aligned}
\end{multline*}
in such a way that:
\Bi
\item there is still the embedding
\begin{multline*}
(\wt M^{{\rm (Bar)};T_p-S_p}_{{\rm ST}_R}\otimes
\wt M^{{\rm (Bar)};T_p-S_p}_{{\rm ST}_L})
\\[8pt]
\subset \quad 
(\wt M^{{\rm (Bar)};T_p-S_p}_{{\rm MG}_R}\otimes
\wt M^{{\rm (Bar)};T_p-S_p}_{{\rm MG}_L})
\quad  \subset \quad 
(\wt M^{{\rm (Bar)};T_p-S_p}_{{\rm M}_R}\otimes
\wt M^{{\rm (Bar)};T_p-S_p}_{{\rm M}_L})
\end{multline*}
between the {\bbf ``~ST~'', ``~MG~'' and ``~M~'' string fields, each one decomposing into:}
\Bean
\item {\bbf a core central time string field;}
\item {\bbf the diagonal, magnetic and electric string fields of the three bisemiquarks;}
\item {\bbf gravitational, magnetic and electric (string) fields of interaction between different right and left semiquarks;}
\item {\bbf strong gravitational and electric (string) fields} of interaction between the \rl core central time string semifield and the \lr space-time semifields of the three semiquarks;
\Ee
as it was developed in proposition 2.15 for the
``~ST~'' shell.

\item The six other ``string'' fields
$(\wt M^{{\rm (Bar)};T_p-S_p}_{{\rm ST}_R}\otimes
\wt M^{{\rm (Bar)};T_p-S_p}_{{\rm MG}_L})$ \dots
$ (\wt M^{{\rm (Bar)};T_p-S_p}_{{\rm M}_R}\otimes
\wt M^{{\rm (Bar)};T_p-S_p}_{{\rm MG}_L})$
are {\bbf interaction (string) fields between different right and left semifields
``~ST~'', ``~MG~'' and ``~M~''.} Each one, for example
$(\wt M^{{\rm (Bar)};T_p-S_p}_{{\rm ST}_R}\otimes
\wt M^{{\rm (Bar)};T_p-S_p}_{{\rm MG}_L})$,
decomposes into:
\Bean
\item an interaction (string field)
$(\wt M^{{\rm (Bar)};T_p}_{{\rm ST}_R}\otimes
\wt M^{{\rm (Bar)};T_p}_{{\rm MG}_L})$ between the right
``~ST~'' core central time string semifield and the left
``~MG~'' core central time string semifield given by the exchange of mixed
``~ST~''-``~MG~'' time quanta;

\item interaction (string) fields
$(\wt M^{q_i;T_p-S_p}_{{\rm ST}_R}\otimes
\wt M^{q_j;T_p-S_p}_{{\rm MG}_L})$, $i=j$ and $i\neq j$, between right ``~ST~'' semifields of right semiquarks and left
``~MG~'' semifields of left semiquarks.

They are:
\Be
\item diagonal (case $i=j$) or gravitational (case $i\neq j$) space and time interaction (string) fields;
\item magnetic interaction (string) fields;
\item electric interaction (string) fields.
\Ee

\item \Bi
\item mixed strong gravitational (string) fields of time type
$(\wt M^{{\rm (Bar)};T_p}_{{\rm ST}_R}\otimes
\wt M^{q_i;T_p}_{{\rm MG}_L})$ between the right
``~ST~'' core central time string semifield and the left
``~MG~'' semifield of time type of the left semiquarks;

\item mixed strong electric (string) fields
$(\wt M^{{\rm (Bar)};T_p}_{{\rm ST}_R}\otimes_e
\wt M^{q_i;S_p}_{{\rm MG}_L})$
between the right
``~ST~'' core central time string semifield and the left
``~MG~'' semifields of space type of the left semiquarks.
\Ei
\Ee
\Ei

\subsubsection{Internal structure of neutrinos}

Assume that a simple massive bisemilepton $e^-$, $\mu ^-$ or $\tau ^-$ is submitted to the inverse deformation morphism:
\begin{multline*}
\Ds^{[M]\to[M\to0]}\RL : \qquad
(\wt M^{T_p-S_p}_{{\rm ST}_R-{\rm MG}_R-{\rm M}_R}\otimes
\wt M^{T_p-S_p}_{{\rm ST}_L-{\rm MG}_L-{\rm M}_L})
\\[8pt]
\To
(\wt M^{T_p-S_p}_{{\rm ST}_R-{\rm MG}_R-({\rm M}_R\to0)}\otimes
\wt M^{T_p-S_p}_{{\rm ST}_L-{\rm MG}_L-({\rm M}_L\to0)})
\end{multline*}
where $\Ds^{[M]\to[M\to0]}\RL$ is the morphism considered in proposition 2.6 responsible for {\bbf the blow-up of the mass shell} in such a way that almost all the mass biquanta become free.  This morphism
$\Ds^{[M]\to[M\to0]}\RL$ is mathematically a (bi)endomorphism based on Galois (bi)antiautomorphisms.  It transforms the bisemilepton $e^-$, $\mu ^-$ or $\tau ^-$ into its corresponding neutrino $\nu _{e^-}$, $\nu _{\mu ^-}$ or $\nu _{\tau ^-}$.  This transformation occurs especially in
$e^++e^- \to \nu _{e^+}+\nu _{e^-}$, i.e. when a bisemielectron splits into a right and a left semielectron in such a way that they are no more localized in a same open ball.  The electric charge of the bisemielectron is then caught by the left or right semielectron, the other semielectron generating its own electric charge by an exchange of electric biquanta between right and left shells of time and space type.

In brief, a (bisemi)neutrino has the same internal structure as the corresponding (bisemi)lepton except that its mass shell tends to zero.
\vskip 11pt

\subsubsection{Possible structure of the dark energy}

In this perspective, {\bbf if the simple and composite bisemifermions are only endowed with their ``~ST~'' internal (vacuum) structure} as considered in propositions 2.14 and 2.15, {\bbf they could correspond to the dark energy} since they are unobservable.  They constitute then a huge reservoir of energy since they are all able to generate their mass shells becoming then observable.
\vskip 11pt

\subsection{Bilinear interactions between \BSMPA}

\subsubsection{Proposition (Structure fields of  $J$ interacting \BSMPA)}

{\em  Let
$(\wt M^{T_p-S_p}_{{\rm ST}_R-{\rm MG}_R-{\rm M}_R}\otimes
\wt M^{T_p-S_p}_{{\rm ST}_L-{\rm MG}_L-{\rm M}_L})$ denote the
``~ST~'', ``~MG~'' and ``~M~'' \BSMST of a massive \bsmpa (essentially a bisermifermion).

Then, the (operator valued) string fields
``~ST~'', ``~MG~'' and ``~M~'' of a set of ``$J$'' interacting \BSMPA are given by:
\begin{multline*}
(\wt M^{T_p-S_p}_{{\rm ST}_R-{\rm MG}_R-{\rm M}_R}(J)\otimes
\wt M^{T_p-S_p}_{{\rm ST}_L-{\rm MG}_L-{\rm M}_L}(J))
\\[8pt]
= 
\bigoplus_{i=1}^J
(\wt M^{T_p-S_p}_{{\rm ST}_R-{\rm MG}_R-{\rm M}_R}(i)\otimes
\wt M^{T_p-S_p}_{{\rm ST}_L-{\rm MG}_L-{\rm M}_L}(i))
\\[8pt]
\bigoplus_{i\neq j=1}^J
(\wt M^{T_p-S_p}_{{\rm ST}_R-{\rm MG}_R-{\rm M}_R}(i)\otimes
\wt M^{T_p-S_p}_{{\rm ST}_L-{\rm MG}_L-{\rm M}_L}(j))
\end{multline*}
where:
\Bean
\item {\bbf the direct sum
$\bigoplus_{i=1}^J(\centerdot (i)\otimes \centerdot (i))$
refers to the sum of the structure fields
``~ST~'', ``~MG~'' and ``~M~'' of a set of ``$J$'' noninteracting (i.e. free) \BSMPA;}

\item {\bbf the mixed sum
$\bigoplus_{i\neq j=1}^J(\centerdot (i)\otimes \centerdot (j))$
refers to the sum of the bilinear interaction fields
``~ST~'', ``~MG~'' and ``~M~'' between right and left semiparticles belonging to different \BSMPA.}
\Ee}

\begin{proof}
\Bena
\item The operator valued string fields
$(\wt M^{T_p-S_p}_{{\rm ST}_R-{\rm MG}_R-{\rm M}_R}\otimes
\wt M^{T_p-S_p}_{{\rm ST}_L-{\rm MG}_L-{\rm M}_L})$
are bisemisheaves of differentiable bifunctions on algebraic bilinear semigroups
$\GL_2(F_{\o v}\times F_v)$ as developed in propositions 2.3 and 2.6.

Let for instance
$(\wt M^{T_p}_{{\rm ST}_R}\otimes
\wt M^{T_p}_{{\rm ST}_L})$ be the time bisemisheaf of one \bsmpa at the ``~ST~'' level: it will then be rewritten according to
$\FRepsp(\GL_2(F_{\o v}\times F_v))^T_{\rm ST}$.

\item On the other hand, the time bisemisheaf at the
``~ST~'' level of a set of ``$J$'' interacting \BSMPA is
$\FRepsp(\GL_{2J}(F_{\o v}\times F_v))^T_{\rm ST}$; it is according to \cite{Pie2} and \cite{Pie3} {\bbf the completely reducible non-orthogonal representation space of
$(\GL_{2J}(F_{\o v}\times F_v))^T_{\rm ST}$ decomposing into:}\pagebreak
\begin{multline*}
\FRepsp(\GL_{2J}(F_{\o v}\times F_v))^T_{\rm ST}
\\[8pt]
\begin{aligned}
&= 
\FRepsp(\GL_{2_1}(F_{\o v}\times F_v))^T_{\rm ST}
\Times \dots 
\begin{aligned}[t]
&\Times
\FRepsp(\GL_{2_i}(F_{\o v}\times F_v))^T_{\rm ST}
\\[8pt]
&\qquad \Times \dots \Times
\FRepsp(\GL_{2_J}(F_{\o v}\times F_v))^T_{\rm ST}
\end{aligned}
\\[8pt]
&= 
\L(
\bigoplus_{i=1}^J\FRepsp(\GL_{2_i}(F_{\o v})^T_{\rm ST}\R)
\times
\L(\bigoplus_{j=1}^J\FRepsp(\GL_{2_j}( F_v)^T_{\rm ST}\R)
\\[8pt]
&=
\bigoplus_{i=1}^J\FRepsp(\GL_{2_i}(F_{\o v}\times F_v))^T_{\rm ST}
\bigoplus_{i\neq j}^J\FRepsp(
T^t_{2_i}(F_{\o v})\times T_{2_j}(F_v)
)^T_{\rm ST}
\\[8pt]
&=
\bigoplus_{i=1}^J
(\wt M^{T_p}_{{\rm ST}_R}(i)\otimes
\wt M^{T_p}_{{\rm ST}_L}(i))
\bigoplus_{i\neq j}^J
(\wt M^{T_p}_{{\rm ST}_R}(i)\otimes
\wt M^{T_p}_{{\rm ST}_L}(j))
\end{aligned}
\end{multline*}
where $\Times$ is the cross binary product \cite{Pie6} leading to a splitting of
$\FRepsp(\GL_{2J}(F_{\o v}\times F_v))^T_{\rm ST}$
into $J$ noninteracting fields
$\FRepsp(\GL_{2_i}(F_{\o v}\times F_v))^T_{\rm ST}$
of $J$ \BSMPA
and into $(J^2-J)$ interacting semifields
$\FRepsp(T_{2_i}^t(F_{\o v})\times T_{2_j}(F_v))^T_{\rm ST}$
of the right and left semiparticles belonging to different
\BSMPA.

\item This completely reducible nonorthogonal representation space
$(\wt M^{T_p}_{{\rm ST}_R}(J)\otimes
\wt M^{T_p}_{{\rm ST}_L}(J))$ of
$\GL_{2J}(F_{\o v}\times F_v))^T_{\rm ST}$
decomposing into:
\[
(\wt M^{T_p}_{{\rm ST}_R}(J)\otimes
\wt M^{T_p}_{{\rm ST}_L}(J))
=\bigoplus_{i=1}^J
(\wt M^{T_p}_{{\rm ST}_R}(i)\otimes
\wt M^{T_p}_{{\rm ST}_L}(i))
	\bigoplus_{i\neq j=1}^J
(\wt M^{T_p}_{{\rm ST}_R}(i)\otimes
\wt M^{T_p}_{{\rm ST}_L}(j))
\]
can be naturally extended to the 
``~ST~'', ``~MG~'' and ``~M~'' fields of $J$ interacting \BSMPA
leading to the thesis.\qedhere
\Ee
\end{proof}
\vskip 11pt

\subsubsection{Proposition}

{\em
{\bbf The (operator valued) string fields of a set of $J$ interacting \BSMPA are given explicitly on the
``~ST~'', ``~MG~'' and ``~M~'' shells by:}
\begin{align*}
&(\wt M^{T_p-S_p}_{{\rm ST}_R-{\rm MG}_R-{\rm M}_R}(J)\otimes
\wt M^{T_p-S_p}_{{\rm ST}_L-{\rm MG}_L-{\rm M}_L}(J))
\\[8pt]
&=
( \wt M^{T_p-S_p}_{{\rm ST}_R}(J)
\oplus
\wt M^{T_p-S_p}_{{\rm MG}_R}(J)
\oplus
\wt M^{T_p-S_p}_{{\rm M}_R}(J) )
\otimes
( \wt M^{T_p-S_p}_{{\rm ST}_L}(J)
\oplus
\wt M^{T_p-S_p}_{{\rm MG}_L}(J)
\oplus
\wt M^{T_p-S_p}_{{\rm M}_L}(J) )\qquad \quad  
\end{align*}

\begin{align*}
=&
\L\{
\L[
\bigoplus_{i=1}^J
(\wt M^{T_p-S_p}_{{\rm ST}_R}(i)\otimes
\wt M^{T_p-S_p}_{{\rm ST}_L}(i))
\bigoplus_{i=1}^J
(\wt M^{T_p-S_p}_{{\rm MG}_R}(i)\otimes
\wt M^{T_p-S_p}_{{\rm MG}_L}(i))
\bigoplus_{i=1}^J
(\wt M^{T_p-S_p}_{{\rm M}_R}(i)\otimes
\wt M^{T_p-S_p}_{{\rm M}_L}(i))
\R]
\R.
\\[8pt]
& \qquad \qquad 
\L[
\bigoplus_{i=1}^J
(\wt M^{T_p-S_p}_{{\rm ST}_R}(i)\otimes
\wt M^{T_p-S_p}_{{\rm MG}_L}(i))
\bigoplus_{i=1}^J
(\wt M^{T_p-S_p}_{{\rm MG}_R}(i)\otimes
\wt M^{T_p-S_p}_{{\rm ST}_L}(i))
\R.
\\[8pt]
& \qquad \qquad \qquad 
\bigoplus_{i=1}^J
(\wt M^{T_p-S_p}_{{\rm ST}_R}(i)\otimes
\wt M^{T_p-S_p}_{{\rm M}_L}(i))
\bigoplus_{i=1}^J
(\wt M^{T_p-S_p}_{{\rm M}_R}(i)\otimes
\wt M^{T_p-S_p}_{{\rm ST}_L}(i))
\\[8pt]
& \qquad \qquad \qquad 
\L.
\L.
\bigoplus_{i=1}^J
(\wt M^{T_p-S_p}_{{\rm MG}_R}(i)\otimes
\wt M^{T_p-S_p}_{{\rm M}_L}(i))
\bigoplus_{i=1}^J
(\wt M^{T_p-S_p}_{{\rm M}_R}(i)\otimes
\wt M^{T_p-S_p}_{{\rm MG}_L}(i))
\R]\R\}
\\[8pt]
&
\begin{aligned}[t]
\;\;\oplus\L\{
\L[
\bigoplus_{i\neq j=1}^J
(\wt M^{T_p-S_p}_{{\rm ST}_R}(i)\otimes
\wt M^{T_p-S_p}_{{\rm ST}_L}(j))
\bigoplus_{i\neq j=1}^J
(\wt M^{T_p-S_p}_{{\rm MG}_R}(i)\otimes
\wt M^{T_p-S_p}_{{\rm MG}_L}(j))
\R.
\R.\qquad 
\\[8pt]
\L.
\L.
\bigoplus_{i\neq j=1}^J
(\wt M^{T_p-S_p}_{{\rm M}_R}(i)\otimes
\wt M^{T_p-S_p}_{{\rm M}_L}(j))
\R]
\R.& \end{aligned}
\\[8pt]
& \qquad \qquad 
\L[
\bigoplus_{i\neq j=1}^J
(\wt M^{T_p-S_p}_{{\rm ST}_R}(i)\otimes
\wt M^{T_p-S_p}_{{\rm MG}_L}(j))
\bigoplus_{i\neq j=1}^J
(\wt M^{T_p-S_p}_{{\rm MG}_R}(i)\otimes
\wt M^{T_p-S_p}_{{\rm ST}_L}(j))
\R.
\\[8pt]
& \qquad \qquad \qquad 
\bigoplus_{i\neq j=1}^J
(\wt M^{T_p-S_p}_{{\rm ST}_R}(i)\otimes
\wt M^{T_p-S_p}_{{\rm M}_L}(j))
\bigoplus_{i\neq j=1}^J
(\wt M^{T_p-S_p}_{{\rm M}_R}(i)\otimes
\wt M^{T_p-S_p}_{{\rm ST}_L}(j))
\\[8pt]
& \qquad \qquad \qquad 
\L.
\L.
\bigoplus_{i\neq j=1}^J
(\wt M^{T_p-S_p}_{{\rm MG}_R}(i)\otimes
\wt M^{T_p-S_p}_{{\rm M}_L}(j))
\bigoplus_{i\neq j=1}^J
(\wt M^{T_p-S_p}_{{\rm M}_R}(i)\otimes
\wt M^{T_p-S_p}_{{\rm MG}_L}(j))
\R]\R\}
\end{align*}
where:
\Bean
\item {\bbf the direct sum of fields inside the first parenthesis $\{\centerdot\}$ refers to the internal structure fields
``~ST~'', ``~MG~'' and ``~M~'' of the $J$ free \BSMPA as well as to the interacting electro-magnetic fields between right and left different shells of these \BSMPA;}

\item {\bbf the direction sum of interaction fields inside the second parenthesis $\{\centerdot\}$ refers to the gravito-electro-magnetic fields of interaction} between the right and left semifields belonging to the same and different shells of the right and left \SMPA dealing with different \BSMPA.
\Ee
}
\vskip 11pt

\begin{proof}
It then appears that the bilinear fields of interaction between $J$ interacting (bisemi)\-particles are very complex if the interactions between the three different shells
``~ST~'', ``~MG~'' and ``~M~'' are taken into account.

More concretely, it appears that:
\Bean
\item in the first parenthesis $\{[\centerdot]\}$, there are six types of diagonal, electro-magnetic fields of interaction between the different shells of the right and left semiparticles of the free \BSMPA, $1\le i\le J$.

Let us consider for example, the interacting fields
$(\wt M^{T_p-S_p}_{{\rm ST}_R}(i)\otimes
\wt M^{T_p-S_p}_{{\rm MG}_L}(i))$ between the right
``~ST~'' shell of the $i$-th right semiparticle and the left
``~MG~'' shell of the symmetric $i$-th left semiparticle of the envisaged $i$-th \bsmpa.

They decompose standardly according to:
\begin{multline*}
(\wt M^{T_p-S_p}_{{\rm ST}_R}(i)\otimes
\wt M^{T_p-S_p}_{{\rm MG}_L}(i))
\\[8pt]
= 
(\wt M^{T_p-S_p}_{{\rm ST}_R}(i)\otimes_D
\wt M^{T_p-S_p}_{{\rm ST}_L}(i))
\oplus
(\wt M^{S_p}_{{\rm ST}_R}(i)\otimes_m
\wt M^{S_p}_{{\rm MG}_L}(i))
\oplus
(\wt M^{(T_p)-S_p}_{{\rm ST}_R}(i)\otimes_e
\wt M^{T_p-(S_p)}_{{\rm MG}_L}(i))
\;.\end{multline*}

\item in the second parenthesis $\{\centerdot\}$, there are also six types of gravitational, electro-magnetic fields of interaction between the shells
``~ST~'', ``~MG~'' and ``~M~'' of the right and left semiparticles belonging to different bisemiparticles, $1\le i,j\le J$, $i\neq j$.

For instance, the gravito-electro-magnetic fields between the right
``~ST~'' shell of the $i$-th right \smpa and the left
 ``~M~'' shell of the $j$-th left \smpa are given by:
 \begin{multline*}
(\wt M^{T_p-S_p}_{{\rm ST}_R}(i)\otimes
\wt M^{T_p-S_p}_{{\rm M}_L}(j))
\\[8pt]
= 
(\wt M^{T_p-S_p}_{{\rm ST}_R}(i)\otimes_D
\wt M^{T_p-S_p}_{{\rm M}_L}(j))
\oplus
(\wt M^{S_p}_{{\rm ST}_R}(i)\otimes_m
\wt M^{S_p}_{{\rm M}_L}(j))
\oplus
(\wt M^{(T_p)-S_p}_{{\rm ST}_R}(i)\otimes_e
\wt M^{T_p-(S_p)}_{{\rm M}_L}(j))
\end{multline*}
where
$(\wt M^{T_p-S_p}_{{\rm ST}_R}(i)\otimes_D
\wt M^{T_p-S_p}_{{\rm M}_L}(J))$
decomposes into ``time'' and ``space'' gravitational fields.

Notice that the electromagnetic fields of interaction between the shells  ``~M~'' of right and left \BSMPA belonging to different \BSMPA corresponds to the exchange of virtual photons in QFT.\qedhere
\Ee
\end{proof}
\vskip 11pt

\subsubsection{Proposition}

{\em 
{\bbf The (operator valued) string fields of a set of $J$ interacting (bisemi)leptons are given on the
``~ST~'', ``~MG~'' and ``~M~'' shells by}
\Bean
\item {\bbf the internal diagonal and electro-magnetic fields of structure between the three shells of these bisemileptons;}

\item {\bbf the gravito-electro-magnetic fields of interaction between left and right semileptons} belonging to different bisemileptons on the same left and right
``~ST~'', ``~MG~'' and ``~M~'' shells;

\item {\bbf the gravito-electro-magnetic fields of interaction between left and right different (i.e. mixed) shells} of left and right semileptons with the hypothesis that these GEM fields in c) are less important than the (G)EM fields considered in a) and in b).
\Ee}
\vskip 11pt

\begin{proof}
In fact, the string fields of a set of $J$ interacting bisemileptons are exactly those considered in proposition 3.2.2.

Taking into account the importance of the (G)EM fields of interaction between the same left and right shells of left and right semileptons, it is reasonable to admit that the GEM fields of interaction between different left and right shells of left and right semileptons are relatively negligible.
\end{proof}
\vskip 11pt

\subsubsection{Proposition}

{\em
{\bbf The (operator valued) string fields of a set of $I$ interacting (bisemi)baryons are given on the
``~ST~'', ``~MG~'' and ``~M~'' shells by:}
\Bean
\item {\bbf the internal string fields of structure between the three shells of these bisemi\-baryons} as developed in section 3.1.5;

\item {\bbf the interaction   gravito-electro-magnetic fields between left and right semi\-baryons belonging to different bisemibaryons} on the same, and different, left and right
``~ST~'', ``~MG~'' and ``~M~'' shells.
\Ee

{\bbf They decompose into \cite{Pie5}:
\Bena 
\item strong gravitational fields between the three shells of the core central time structures belonging to different left and right semibaryons;

\item gravito-electro-magnetic fields between the three shells of left and right semiquarks belonging to different semibaryons;

\item strong gravitational and electric fields between the three shells of \rl core central time structures and \lr semiquarks belonging to different semibaryons.
\Ee}

They are responsible for the generation of mesons, as it was developed in \cite{Pie5}.
}
\vskip 11pt

\begin{proof}[Sketch of proof]

The GEM fields of interaction between nonsymmetric left and right semibaryons differ from the corresponding GEM fields of interaction between nonsymmetric left and right semileptons by the existence of strong gravitational fields between the left and right shells of core central time semistructures (case 1)) and strong gravitational and electric fields (case 3)) at the origin of mesons.

These strong fields of interaction are due to the existence of core central time structures of semibaryons.

As in the leptonic case, it is reasonable to admit that the GEM fields of interaction between different left and right shells of left and right semibaryons are relatively negligible.
\end{proof}
\vskip 11pt

\subsubsection{Proposition}

{\em
{\bbf The interaction (string) fields between a set of $I$ bisemibaryons and a set of $J$ bisemileptons are given on the
``~ST~'', ``~MG~'' and ``~M~'' shells by:
\Bean
\item gravitational and electric fields between core time structures of \rl semibaryons and space-time structures of \lr semileptons;

\item gravito-electro-magnetic fields between space-time structures of right \resp{\linebreak left} semiquarks and space-time structures of \lr semileptons.
\Ee}}
\vskip 11pt

\begin{proof}
The (operator values) string fields of a set of $J$ bisemileptons interacting with a set of $I$ bisemibaryons are given on the set
``~ST$\oplus$MG$\oplus$M~'' levels (noted ``~ST--MG--M~'') according to proposition 3.2.1 by:
\begin{multline*}
\L[
\bigoplus_{i=1}^J 
\wt M^{T_p-S_p}_{{\rm ST-MG-M}_R}(\ell_i)
\bigoplus_{k=1}^I
\wt M^{{\rm (Bar)};T_p-S_p}_{{\rm ST-MG-M}_R}(b_k )
\R]
\otimes
\L[
\bigoplus_{j=1}^J 
\wt M^{T_p-S_p}_{{\rm ST-MG-M}_L}(\ell_j)
\bigoplus_{\ell=1}^I
\wt M^{{\rm (Bar)};T_p-S_p}_{{\rm ST-MG-M}_L}(b_\ell )
\R]
\\[8pt]
\begin{aligned}[t]
&= 
\bigoplus_{i,j=1}^J 
( \wt M^{T_p-S_p}_{{\rm ST-MG-M}_R}(\ell_i)
\otimes
\wt M^{T_p-S_p}_{{\rm ST-MG-M}_L}(\ell_j))
\\[8pt]
&\qquad  
\bigoplus_{k,\ell=1}^I 
( \wt M^{{\rm (Bar)};T_p-S_p}_{{\rm ST-MG-M}_R}(b_k)
\otimes
\wt M^{{\rm (Bar)};T_p-S_p}_{{\rm ST-MG-M}_L}(b_\ell))
\\[8pt]
&\qquad  
\bigoplus_{i=1}^J 
\bigoplus_{\ell=1}^I 
( \wt M^{T_p-S_p}_{{\rm ST-MG-M}_R}(\ell _i)
\otimes
\wt M^{{\rm (Bar)};T_p-S_p}_{{\rm ST-MG-M}_L}(b_\ell))
\\[8pt]
&\qquad  
\bigoplus_{k=1}^I
\bigoplus_{j=1}^J 
( \wt M^{{\rm (Bar)}; T_p-S_p}_{{\rm ST-MG-M}_R}(b_k)
\otimes
\wt M^{T_p-S_p}_{{\rm ST-MG-M}_L}(\ell_j))
\end{aligned}
\end{multline*}
where:
\Bean
\item $\bigoplus_{i,j=1}^J 
( \wt M^{T_p-S_p}_{{\rm ST-MG-M}_R}(\ell_i)
\otimes
\wt M^{T_p-S_p}_{{\rm ST-MG-M}_L}(\ell_j))$
refers to the string fields of a set of $J$ interacting bisemileptons as developed in proposition 3.2.2 and 3.2.3.

\item $\bigoplus_{k,\ell=1}^I 
( \wt M^{{\rm (Bar)};T_p-S_p}_{{\rm ST-MG-M}_R}(b_k)
\otimes
\wt M^{{\rm (Bar)};T_p-S_p}_{{\rm ST-MG-M}_L}(b_\ell))$
refers to the string fields of a set of $I$ interacting bisemibaryons as developed in proposition 3.2.4.

\item $\bigoplus_{i=1}^J
\bigoplus_{\ell=1}^I
( \wt M^{ T_p-S_p}_{{\rm ST-MG-M}_R}(\ell _i)
\otimes
\wt M^{{\rm (Bar)};T_p-S_p}_{{\rm ST-MG-M}_L}(b_\ell ))$
(and similarly,
$\bigoplus_{k=1}^I
\bigoplus_{j=1}^J 
( \wt M^{{\rm (Bar)}; T_p-S_p}_{{\rm ST-MG-M}_R}(b_k)
\otimes
\wt M^{T_p-S_p}_{{\rm ST-MG-M}_L}(\ell_j))$)
refers to the interaction fields between a set of $J$ \rl semileptons and a set of $I$ \lr semibaryons.

They decompose according to:\pagebreak
\begin{multline*}
\bigoplus_{i=1}^J
\bigoplus_{\ell=1}^I
( \wt M^{ T_p-S_p}_{{\rm ST-MG-M}_R}(\ell _i)
\otimes
\wt M^{{\rm (Bar)};T_p-S_p}_{{\rm ST-MG-M}_L}(b_\ell ))
\\[8pt]
\begin{aligned}[t]
&=
\bigoplus_{i=1}^J
\bigoplus_{\ell=1}^I
( \wt M^{ T_p-S_p}_{{\rm ST-MG-M}_R}(\ell _i)
\otimes
(\wt M^{{\rm (Bar)};T_p(r)}_{{\rm ST-MG-M}_L}(b_\ell )
\bigoplus_{\alpha =1}^3
\wt M^{q_{\ell _\alpha };T_p-S_p}_{{\rm ST-MG-M}_L}(b_\ell )))
\\[8pt]
&=
\bigoplus_{i=1}^J
\bigoplus_{\ell=1}^I
\L[
( \wt M^{ T_p-S_p}_{{\rm ST-MG-M}_R}(\ell _i)
\otimes
\wt M^{{\rm (Bar)};T_p(r)}_{{\rm ST-MG-M}_L}(b_\ell ))\R.
\\[8pt]
& \qquad \qquad 
\L.
\bigoplus_{\alpha =1}^3
(\wt M^{ T_p-S_p}_{{\rm ST-MG-M}_R}(\ell _i)
\otimes
(\wt M^{q_{\ell _\alpha };T_p-S_p}_{{\rm ST-MG-M}_L}(b_\ell ))\R]
\end{aligned}
\end{multline*}
where:
\Be
\item $( \wt M^{ T_p-S_p}_{{\rm ST-MG-M}_R}(\ell _i)
\otimes
\wt M^{{\rm (Bar)};T_p(r)}_{{\rm ST-MG-M}_L}(b_\ell ))$
\\[8pt]
$
= 
( \wt M^{ T_p}_{{\rm ST-MG-M}_R}(\ell _i)
\otimes_D
\wt M^{{\rm (Bar)};T_p(r)}_{{\rm ST-MG-M}_L}(b_\ell ))
\oplus
( \wt M^{S_p}_{{\rm ST-MG-M}_R}(\ell _i)
\otimes_e
\wt M^{{\rm (Bar)};T_p(r)}_{{\rm ST-MG-M}_L}(b_\ell ))$
\\[8pt]
decomposes into a (time) gravitational field
$( \wt M^{ T_p}_{{\rm ST-MG-M}_R}(\ell _i)
\otimes_D
\wt M^{{\rm (Bar)};T_p(r)}_{{\rm ST-MG-M}_L}(b_\ell ))$ which may be strong (see corollary 3.2.6) and into an electric field
$( \wt M^{S_p}_{{\rm ST-MG-M}_R}(\ell _i)
\otimes_e
\wt M^{{\rm (Bar)};T_p(r)}_{{\rm ST-MG-M}_L}(b_\ell ))$
which may also be strong;

\item $\bigoplus_{\alpha =1}^3
(\wt M^{ T_p-S_p}_{{\rm ST-MG-M}_R}(\ell _i)
\otimes
\wt M^{q_{\ell _\alpha };T_p-S_p}_{{\rm ST-MG-M}_L}(b_\ell ))$
generate mixed gravito-electro-magnetic\linebreak fields of interaction between the right semilepton $\ell _i$ and the three left semiquarks
$q_{\ell _\alpha }$ of the left semibaryon $b_\ell $ (see, for instance, proposition 2.15).\qedhere
\Ee
\Ee
\end{proof}
\vskip 11pt

\subsubsection{Corollary}

{\em In the case of deep inelastic scattering of leptons by baryons, the gravitational and electric fields between core time structures of semibaryons and space-time structures of semileptons can be at the origin of the production of (bisemi)hadrons.
}
\vskip 11pt

\begin{proof}
If we consider, for example, the deep inelastic scattering of electrons by protons, then the gravitational and electric fields between core time structures of semibaryons and space-time structures of semileptons are strong according to proposition 3.2.5 and they can generate massive (bisemi)baryons and mesons as it was seen in proposition 2.16 and in \cite{Pie5}.
\end{proof}
\vskip 11pt

The developments of chapters 2 and 3 lead us then to the following.
\vskip 11pt

\subsubsection{Main proposition ((Strong) GEM forces)}

{\em
{\bbf The four fundamental forces inside and between massive (bisemi)fermions result from bilinear interactions between right and left semifermions on the three embedded shells
``~ST~'', ``~MG~'' and ``~M~''.

They are essentially gravito-electro-magnetic forces on the set of these three embedded shells.

They can be classified according to:}
\Bena
\item electromagnetic forces inside elementary (i.e. simple) bisemifermions.  They are mediated by massless EM bisemibosons;

\item gravito-electro-magnetic forces inside composite bisemifermions, i.e. bisemibaryons.  They are mediated by the massless GEM bisemibosons;

\item strong gravitational and electric forces inside (perturbed) bisemibaryons.  They are mediated by massive bisemibosons, or mesons;

\item gravito-electro-magnetic forces between interacting bisemifermions.  They are mediated by massless GEM bisemibosons;

\item strong gravitational and electric forces between interacting perturbed bisemibaryons.  They are at the origin of the classical strong and weak forces and are mediated by mesons, i.e. massive bisemibosons.
\Ee}
\vskip 11pt

\begin{proof}
The classification of forces inside and between (massive) bisemifermions are essentially gravito-electro-magnetic forces as it was seen in chapter 3.  More specifically, we have that:
\Bena
\item Inside an elementary (massive) bisemifermion, i.e. a bisemilepton or a bisemiquark, there are, on each 
``~ST~'', ``~MG~'' and ``~M~'' shell, an internal magnetic field and an internal electric field responsible for the electric charge as developed in lemma 3.1.1, in proposition 2.14 and in section 3.1.4;

\item Inside a massive bisemibaryon, there are, on each
``~ST~'', ``~MG~'' and ``~M~'' shell:
\Be
\item magnetic and electric fields inside the three bisemiquarks;

\item gravitational, magnetic and electric fields of interaction between different right and left semiquarks;

\item strong gravitational and electric fields of interaction between the \rl core central time semifield and the \lr space-time semifields of the three semiquarks;
\Ee
as shown in section 3.1.5 and in proposition 2.15;

\item These strong gravitational and electric fields of 2)c) inside a perturbed bisemibaryon can be at the origin of the generation of massive mesons as developed in proposition 2.16;

\item A set of massive bisemileptons interacts by means of gravito-electro-magnetic fields mediated by massless GEM bisemibosons on the three embedded shells
``~ST~'', ``~MG~'' and ``~M~'' as proved in proposition 3.2.3;

\item A set of massive (perturbed) bisemibaryons interacts on three shells
``~ST~'', ``~MG~'' and ``~M~'' by means of:
\Be
\item strong gravitational fields between left and right core central time structures of different bisemibaryons;

\item gravito-electro-magnetic fields between left and right semiquarks belonging to different bisemibaryons;

\item strong gravitational and electric fields between \rl core central time structures and \lr semiquarks belonging to different bisemibaryons.

They can be at the origin of the generation of massive mesons.
\Ee
\Ee

This results from proposition 3.2.4.
\end{proof}
\vskip 11pt

\subsubsection{Proposition (Mixed fundamental forces)}

{\em
Besides the four fundamental forces on the products, right by left, ${\rm ST}_R\times {\rm ST}_L$, ${\rm MG}_R\times {\rm MG}_L$ and ${\rm M}_R\times {\rm M}_L$ of the space-time, middle ground and mass shells, {\bbf there exist mixed fundamental forces, which are (strong) gravito-electro-magnetic forces, between:
\Bean
\item the space-time and middle ground levels;
\item the space-time and mass levels;
\item the middle ground and mass levels.
\Ee
These mixed fundamental forces} between different left and right shells {\bbf are supposed to be small} with respect to the fundamental forces between same left and right shells.
}
\vskip 11pt

\begin{proof}
Referring to section 3.1.4 describing the internal structure of a simple massive bisemi\-fermion, we see that there exist six mixed fields, which are electro-magnetic fields, between different left and right ST~, MG and M shells.

Generalizing this fact to the internal structure of a bisemibaryon and to sets of interacting bisemileptons and bisemibaryons, we see easily that all the four fundamental forces considered in proposition 3.2.7 between the same right and left shells can be transposed to different right and left shells ST~, MG and M~.

So, these four mixed fundamental forces are (strong) gravito-electro-magnetic forces which are supposed to be small with respect to the fundamental forces on the same shells due to an effect of congestion of exchanged biquanta and to the different natures (and rotational velocities) of these shells.
\end{proof}
\vskip 11pt

\subsubsection{The weak decays}

\Bi
\item The decays considered until now in this paper are {\bbf nonleptonic decays giving rise to the production of massive mesons from strong gravitational and electric fields} inside (or between) bisemibaryons (see, for instance, proposition 2.16, section3.1.5 (and proposition 3.2.4)).

\item {\bbf The leptonic decays} of the form
\[ A\to B+\ell +\nu _\ell  \qquad
\Longrightarrow \quad
q_i\to q_{i'}+\ell +\nu _\ell \;, \qquad q_i\in A\; , \quad q_{i'}\in B\;, \]
of a bisemibaryon $A$ results from the diagonal emission of a bisemilepton $\ell $ by a bisemiquark $q_i$ of $A$ throughout a biendomorphism of $q_i$ transforming it into a lighter bisemiquark $q_{i'}$ of a bisemibaryons $B$ in such a way that a bisemineutrino $\nu _\ell $ is emitted to take into account the bilinear interaction between the bisemiquark $q_{i'}$ and the bisemilepton $\ell $.
\Ei
\vskip 11pt

This was developed in proposition 5.4.2 of \cite{Pie1} and will be succinctly recalled in the following proposition.
\vskip 11pt

\subsubsection{Proposition (Leptonic decays)}

{\em
{\bbf The emission of a bisemineutrino in a leptonic decay results from the condensation of the GEM field of interaction between the bisemiquark $q_{i'}$ of the generated bisemibaryon $B$ and the emitted massive bisemilepton $\ell $.
}}
\vskip 11pt

\begin{proof}
Let $A\to B+\ell +\nu _\ell $ such that $q_i\to q_{i'}+
\ell +\nu _\ell $ 
be the leptonic decay envisaged in section~3.2.9.

Let $(\wt M^{q_i;T_p-S_p}_{{\rm ST-MG-M}_R}\otimes
\wt M^{q_i;T_p-S_p}_{{\rm ST-MG-M}_L})$ be the
ST~, MG and M shells of the bisemiquark $q_i\in A$.

Let $(E^D_R\otimes_D E^D_L)$ be the diagonal smooth biendomorphism applied to the diagonal structures of the bisemiquak $q_i$:
\begin{multline*}
(E^D_R\otimes_D E^D_L): \qquad
(\wt M^{q_i;T_p-S_p}_{{\rm ST-MG-M}_R}\otimes_D
\wt M^{q_i;T_p-S_p}_{{\rm ST-MG-M}_L})
\\[8pt]
\To
(\wt M^{q_{i'};T_p-S_p}_{{\rm ST-MG-M}_R}\otimes
\wt M^{q_{i'};T_p-S_p}_{{\rm ST-MG-M}_L})
\oplus
(\wt M^{\ell ;T_p-S_p}_{{\rm ST-MG-M}_R}\otimes
\wt M^{\ell ;T_p-S_p}_{{\rm ST-MG-M}_L})
\\[8pt]
\oplus \L[
(\wt M^{q_{i'};T_p-S_p}_{{\rm ST-MG-M}_R}\otimes
\wt M^{\ell ;T_p-S_p}_{{\rm ST-MG-M}_L})
+
(\wt M^{\ell ;T_p-S_p}_{{\rm ST-MG-M}_R}\otimes
\wt M^{q_{i'};T_p-S_p}_{{\rm ST-MG-M}_L})
\R]
\end{multline*}
in such a way that:
\begin{multline*}
(\wt M^{q_{i};T_p-S_p}_{{\rm ST-MG-M}_R}\otimes
\wt M^{q_{i};T_p-S_p}_{{\rm ST-MG-M}_L})
\\[8pt]
=
\L(
(\wt M^{q_{i'} ;T_p-S_p}_{{\rm ST-MG-M}_R}\oplus
\wt M^{\ell ;T_p-S_p}_{{\rm ST-MG-M}_R}
)
\otimes
(\wt M^{q_{i'};T_p-S_p}_{{\rm ST-MG-M}_R}\oplus
\wt M^{\ell ;T_p-S_p}_{{\rm ST-MG-M}_L})
\R)
\end{multline*}
where:
\Bean
\item $(\wt M^{q_{i'};T_p-S_p}_{{\rm ST-MG-M}_R}\otimes
\wt M^{q_{i'};T_p-S_p}_{{\rm ST-MG-M}_L})
$ are the ST~, MG and M shells of the lighter bisemiquark $q_{i'}
$;

\item $
(\wt M^{\ell ;T_p-S_p}_{{\rm ST-MG-M}_R}\otimes
\wt M^{\ell ;T_p-S_p}_{{\rm ST-MG-M}_L})
$ are the ST~, MG and M shells of the emitted bisemilepton whose shell structures are the difference between the $q_i$ and $q_{i'}$ bisemiquark shell structures in the sense of proposition 5.4.2 of \cite{Pie1};

\item $
(\wt M^{q_{i'};T_p-S_p}_{{\rm ST-MG-M}_R}\otimes
\wt M^{\ell ;T_p-S_p}_{{\rm ST-MG-M}_L})
\oplus
(\wt M^{\ell ;T_p-S_p}_{{\rm ST-MG-M}_R}\otimes
\wt M^{q_{i'};T_p-S_p}_{{\rm ST-MG-M}_L})$
can be decomposed into gravito-electro-magnetic fields of interaction between the three shells of the bisemiquarks $q_{i'}$ and the bisemilepton $\ell $ and are able to condense into the sum
$\nu _\ell =\mu _\ell +\mu _{q_{i'}}$ of two light bisemileptons $\mu _\ell $ and $\mu _{q_{i'}}$.

This new light bisemilepton of interaction $\nu _\ell $ is interpreted as the neutrino of the bisemilepton $\ell $.
\qedhere
\Ee
\end{proof}
\vskip 11pt

\subsubsection{Emitted mesons and neutrinos}

Remark that {\bbf the emitted mesons} in the strong weak decays
{\bbf result from the condensation of strong gravitational and electric fields} according to proposition 2.16, section 3.1.5 and proposition 3.2.4 while {\bbf the emitted (bisemi)neutrinos} in the (leptonic) weak decays {\bbf result from the condensation of gravito-electro-magnetic fields} of interaction {\bbf which are not strong} since the core central time structures are assumed not to be involved in this last case.


\begin{thebibliography}{99}
\addcontentsline{toc}{section}{References}

\newcommand\auteur[2]{\bibitem[#1]{#1} {\sc #2,\/}}
\newcommand\titre[1]{#1,}
\newcommand\revue[4]{{\em #1\/} {\bf #2} (#3), #4.}
\newcommand\book[3]{{\em #1\/}, (#2), {#3}}

\auteur{A-L}{E. Abers, B. Lee} 
\titre{Gauge theories} 
\revue{Phys. Rep.}{C9} {1973} {1--141}

\auteur{A-J} {M. Atiyah, J. Jones} 
\titre{Topological aspects of Yang-Mills theory}
\revue{Commun. Math. Phys.} {61} {1978} {97--118}

\auteur{Ash} {A. Ashtekar} 
\titre{New variables for classical and quantum gravity}
\revue{Phys. Rev. Lett.}{57}{1986}{2244--2247}

\auteur{Ati}{M. Atiyah}
\titre{Reflections on geometry and physics}
\revue{Surveys in Diff. Geom.}{2}{1995}{1--6}

\auteur{B-E}{R. Brout, F. Englert}
\titre{Broken symmetry and the mass of gauge vector mesons}
\revue{Phys. Rev. Lett.}{13}{1964}{321}

\auteur{Bou}{N. Bourbaki}
\book{Algèbre Commutative. Ch. 1 \& 2}{Hermann}{1961.}

\auteur{Car}{H. Carayol}
\titre{Preuve de la conjecture de Langlands locale pour $\GL_n$, Travaux de Harris-Taylor et Henniart}
\revue{Sém. Bourbaki}{857}{1999}{1--52}

\auteur{Dir}{P.A.M. Dirac}
\titre{The quantum theory of electrons}
\revue{Proc. Soc. London}{A117}{1928}{610--624}

\bibitem[Del$\to$Wit]{DelToWit} {\sc P. Deligne, P. Etingov, D. Freed, L. Jeffrey, D. Kazhdan, J. Morgan, D. Morrison, E. Witten,}  
Quantum fields and strings: a course for mathematicians. Vols. 1 \& 2. 
{\em Amer. Math. Soc.\/} and {\em Inst. Adv. Stud.\/} (1999).

\bibitem[Dys]{Dys} {\sc J. Dyson,\/} 
The radiation theory of Tomonaga, Schwinger and Feynman.  
{\em Phys. Rev.\/}, {\bf 75\/} (1949), 486--502.

\bibitem[Ein]{Ein}  {\sc A. Einstein,} 
The principle of relativity. 
Dover (1923).


\auteur{F-L-S}{R. Feynman, R. Leighton, M. Sands}
\book{The Feynman Lectures on Physics, T. II}{Addison-Wesley}{1966.}

\auteur{Gel}{S. Gelbart}
\titre{An elementary introduction to the Langlands program}
\revue{Bull. Amer. Math. Soc.}{10}{1984}{177--219}

\auteur{G-G-S}{M. Gaillard, P. Grannis, F. Sciulli}
\titre{The standard model of particle physics}
\revue{Rev. Mod. Phys.}{71}{1999}{S96--S111}

\auteur{G-S-W}{M. Green, J.M. Schwarz, E. Witten}
\book{Superstring Theory, Vol. 1 \& 2}{Cambridge University Press}{1987.}

\bibitem[G-W]{G-W} {\sc D. Gross, F. Wilckzek,} 
Ultraviolet behavior of non abelian gauge theories. 
{\em Phys. Rev. Letters\/}, {\bf 30\/} (1973), 1343--1346.

\bibitem[Har]{Har} {\sc G. Harder,} 
Eisenstein cohomology of arithmetic groups.  The case $\GL_2$~. 
{\em Invent. Math.\/}, {\bf 89\/} (1987), 37--118.

\auteur{Hig}{P. Higgs}
\titre{Broken symmetries and the masses of gauge bosons}
\revue{Phys. Rev. Lett.}{13}{1964}{508}

\auteur{K-L}{A. Klein, B.W. Lee}
\titre{Does the spontaneous breakdown of symmetry imply zero-mass particles?}
\revue{Phys. Rev. Lett.}{12}{1964}{266}

\bibitem[Kok]{Kok} {\sc J.J. Kokkedee,} 
The quark model.  
Benjamin (1969).

\auteur{Kot}{D. Kotschick}
\titre{Gauge theory is dead!...  Long live gauge theory}
\revue{Notices AMS}{42}{1995}{335--338}

\bibitem[Lang]{Lang} {\sc S. Lang,} 
Introduction to Modular Forms. 
Springer (1976).

\auteur{Lan1}{R. Langlands}
\titre{On the notion of an automorphic representation}
\revue{Proc. Symp. Pure Math.}{33}{1979}{203--208}

\auteur{Lan2}{R. Langlands}
\titre{Where stands functoriality today?}
\revue{Proc. Symp. Pure Math.}{63}{1997}{457--471}

\auteur{Mum}{D. Mumford}
\titre{The red book of varieties and schemes}
\revue{Lect. Not. Math.}{1358}{1988}{Springer}

\bibitem[Pie1]{Pie1} {\sc C. Pierre,} Algebraic quantum theory. 
ArXiv:math-ph/0404024 (2004).

\bibitem[Pie2]{Pie2} {\sc C. Pierre,}  
$n$-dimensional global correspondences of Langlands.  
ArXiv-math:RT/0510348 (2005).

\bibitem[Pie3]{Pie3} {\sc C. Pierre,} 
A new track for unifying general relativity with quantum field theories.  
ArXiv:gr-qc/0510091 (2005).

\bibitem[Pie4]{Pie4} {\sc C. Pierre,}
Brane and string field structure of elementary particles.
ArXiv physics/0607249 (2006).

\bibitem[Pie5]{Pie5} {\sc C. Pierre,}
{Quantized gravito-electro-magnetic interactions of bilinear type}.
{ArXiv physics/0607235}{ (2006)}.

\bibitem[Pie6]{Pie6} {\sc C. Pierre,}
{Introducing bisemistuctures}.
{ArXiv Math GM/0607624 (2006).}

\bibitem[Pie7]{Pie7} {\sc C. Pierre,}
{From global class field concepts and modular representations to the conjectures of STW, BSD and Riemann}.
ArXiv Math RT/0608084 (2006).

\auteur{Pol}{J. Polchinsky}
\book{String theory, Vol. 1 \& 2}{Cambridge Univ. Press}{1998.}

\auteur{Rib}{K. Ribet}
\titre{Galois representations and modular forms}
\revue{Bull. Amer. Math. Soc.}{32}{1995}{375--402}

\auteur{Smo}{L. Smolin}
\titre{On the nature of quantum fluctuations and their relations to gravitation and the principle of inertia}
\revue{Class. Quant. Gravit.}{3}{1986}{347--359}

\auteur{S-S}{J. Schwarz, N. Seiberg}
\titre{String theory, supersymmetry, unification and all that}
\revue{Rev. Mod. Phys.}{71}{1999}{S111--S120}

\auteur{Vneu}{J. Von Neumann}
\book{Mathematical Foundations of Quantum Mechanics}
{Princeton Univ. Press}{1955.}

\auteur{Wei}{S. Weinberg}
\book{The Quantum Theory of Fields, Vol. 1 \& 2}{Cambridge Univ. Press}{1996.}

\auteur{Wil}{F. Wilczek}
\titre{Quantum field theory}
\revue{rev. Mod. Phys.}{71}{1999}{S85--S95}

\auteur{Wit}{E. Witten}
\titre{Reflections on the fate of space-time}
\revue{Physics Today}{april}{1996}{24--30}


\vspace{2cm}
\hfill\begin{minipage}{6cm}
C. Pierre\\
Université de Louvain\\
Chemin du Cyclotron, 2\\
B-1348 Louvain-la-Neuve,  Belgium\\
pierre@math.ucl.ac.be
\end{minipage}

\end{thebibliography}
\end{document}